\documentclass[useAMS,usenatbib]{mn2e_astroph}

\usepackage{graphicx}

\def\nct#1{\nocite{#1}}
\newcommand{\rlc}{R_{\rm lc}}

\def \mref#1{(\ref{#1})} 
\newcommand\sss{\scriptscriptstyle}

\def\npsi{N_{\psi,in}}
\def\wpsi{W_{\psi,in}}
\def\nlag{N_{\Delta\phi}}
\def\wlag{W_{\Delta\phi}}
\def\psin{\psi_{in}}
\def\lpk{\Delta\phi_{pk}}
\def\ppk{\psi_{pk}}
\def\lag{\Delta\phi}

\title[The origin of radio pulsar polarisation]
{The origin of radio pulsar polarisation
}

\author[J.~Dyks]
{J.~Dyks
\\
Nicolaus Copernicus Astronomical Center, Rabia\'nska 8, 87-100, Toru\'n,
Poland\\
}
\begin{document}

\date{Accepted .... Received ...; in original form 2017 May 17}


\maketitle

\label{firstpage}

\begin{abstract}
Polarisation of radio pulsar profiles involves a number of poorly
understood, intriguing   
phenomena, such as the existence of comparable amounts of orthogonal
polarisation modes (OPMs), strong 
distortions of polarisation angle (PA) curves into shapes inconsistent with the
rotating vector model (RVM), 
and the strong circular polarisation $V$ which 
can be maximum (instead of zero) at the OPM jumps. It is shown that 
the existence of comparable OPMs and of the large $V$ results from
a coherent addition of phase-delayed waves in natural propagation modes, 
which are produced when a linearly polarised 
emitted signal propagates through an intervening medium on its way to reach the
observer. The longitude-dependent flux ratio of two OPMs 
can be understood as the result of backlighting the intervening polarisation
basis by the emitted radiation.  
The coherent mode summation implies opposite polarisation 
properties to those known from the incoherent case, 
in particular, the OPM jumps occur at peaks of $V$, whereas $V$ changes sign 
at a maximum of the linear polarisation fraction $L/I$. 
These features are indispensable to interpret various 
observed polarisation effects.
It is shown that statistical properties of the emission mechanism and of 
propagation effects can be efficiently parametrised in a simple 
model of coherent mode addition, 
which is
successfully applied to complex polarisation phenomena, such as the
stepwise PA curve of PSR B1913$+$16 and the strong distortions of the PA 
curve within core components of pulsars B1933$+$16 and B1237$+$25. 
The inclusion of coherent mode addition opens the possibility for 
a number of new polarisation effects, 
such as inversion of relative modal strength, twin minima in $L/I$
coincident with peaks in $V$, $45^\circ$ PA jumps 
in weakly polarised emission, and 
 loop-shaped core PA distortions.
  The empirical treatment of the coherency 
of mode addition 
makes it possible 
to advance the understanding of pulsar polarisation beyond the RVM
model. 
\end{abstract}


\def\lap{\hbox{\hspace{4.3mm}}
         \raise1.5pt \vbox{\moveleft9pt\hbox{$<$}}
         \lower1.5pt \vbox{\moveleft9pt\hbox{$\sim$ }}
         \hbox{\hskip 0.02mm}}

\def\rwobs{R_W}
\def\rwcon{R_W}
\def\rwstr{R_W}
\def\winobs{W_{\rm in}}
\def\woutobs{W_{\rm out}}
\def\phm{\phi_m}
\def\phmi{\phi_{m, i}}
\def\thm{\theta_m}
\def\dres{\Delta\phi_{\rm res}}
\def\win{W_{\rm in}}
\def\wout{W_{\rm out}}
\def\rin{\rho_{\rm in}}
\def\rout{\rho_{\rm out}}
\def\phin{\phi_{\rm in}}
\def\phout{\phi_{\rm out}}
\def\xin{x_{\rm in}}
\def\xout{x_{\rm out}}

\def\thmin{\theta_{\rm min}^{\thinspace m}}
\def\thmax{\theta_{\rm max}^{\thinspace m}}

\section{Introduction}
\label{intro}

The origin of orthogonal polarisation modes (OPMs) in pulsars has been unknown, 
except for  
the general notion that they are probably propagation effects in
 birefringent pulsar plasma. The observed properties of the modes 
make up for a real panopticon of peculiarities, which is well documented 
in pulsar literature.\footnote{For example in Xilouris et al.~(1998),  
Stairs et al.~(1999),  
Weltevrede \& Johnston (2008), Gould \& Lyne (1998), Manchester \& Han
(2004),  
Tiburzi et al.~(2013), Karastergiou \& Johnston (2006), 
Johnston \& Weisberg (2006), 
 Keith et al.~(2009).
\nct{xkj98, stc99, wj08, gl98, mh04, tjb13, kj06, jw06}
} 
It is possible to find polarisation angle (PA) curves which follow the
rotating vector model (RVM), e.g.~B0301$+$19, B0525$+$21 (Hankins \& Rankin
2010, hereafter HR10). \nct{hr10} 
Other PA curves follow RVM, but with frequent transitions (jumps) 
between the modes. Other cases seemingly 
do not obey any simple model (erratic, e.g.~B1946$+$35, Mitra \& Rankin 2017). 
\nct{mr2017}
The PA distributions of single samples (recorded in single pulse
observations) reveal that the polarisation modes can be very well defined 
by narrow peaks, or oppositely, be very wide, almost merging with
each other, or filling the whole $180^\circ$ PA range (Stinebring et
al.~1984, Mitra et al.~2015, hereafter MAR15). 
\nct{scr84, mar2015}
In the latter case the associated linear polarisation 
fraction $L/I$ is very low (eg.~B2110$+$27, Fig.~19 in MAR15), 
but in other cases $L/I$ may be very high.

This complex picture is spiced up with cases which generally follow RVM, but
locally, especially at the central component,   
exhibit extremely
complicated non-RVM distortions (e.g.~B1933$+$16, Mitra et al.~2016, hereafter
MRA16; B1237$+$25, Smith et al.~2013, hereafter SRM13).
\nct{mra2016, srm13}
Several objects, including those with complex core 
polarisation, reveal curious symmetry of their single-pulse PA tracks: 
the strongest (primary) polarisation mode is acompanied by  
short patches of the secondary mode in the profile peripheries. 

The scant attempts to understand the complex polarisation distortions
 involved both coherent and incoherent summation 
of radiation in different modes. 
The coherent interaction of modes has long been considered a possible
source of observed circular polarisation (Cheng \& Ruderman 1979; 
Melrose 2003; Lyubarskii \& Petrova 1998), 
\nct{cr79, lp98, m03}
since the latter appears
when the modal waves combine with a non-zero phase lag (which is not a
multiple of $\pi$). However, propagation processes were also considered unlikely 
to induce appropriately small phase delays (of the order of the wavelength) 
to avoid complete cancelling of $V$ (Michel 1991, p.~30).\nct{m91}
Moreover, the appreciable amount of $V$ can also be directly 
produced by the emission mechanism (e.g.~Sokolov \& Ternov 1968; Gangadhara
2010).
\nct{st68, g10}

The influence of magnetised plasma on propagating waves 
is known (e.g.~Barnard \& Arons, 1986; Wang et al.~2010; 
Beskin and Philippov 2012), 
\nct{ba86, bp12, wlh10}
so the properties of the
outgoing pulsar radiation can in principle be estimated. 
However, the result is sensitive to the poorly known and likely 
complex properties of the intervening medium 
(propagation direction and the spatial distribution of plasma density) 
which makes the detailed modelling 
of propagation physics ambiguous. The complexity of such calculations 
(e.g.~Wang et al.~2010), and the vague correspondence between their results 
\nct{wlh10}
and data offers limited practical help in interpreting 
the observed polarisation.

The alternative empirical approach attempts to reproduce the
observed non-RVM effects through some specific combination of radiation 
in each polarisation mode. This is typically a data-guided guess, also
suffering from the problem of ambiguity. 
The central profile region of PSR B0329$+$54, for example, 
exhibits the elliptically polarised radiation 
with the PA which deviates from the RVM. Edwards \& Stappers (2004) interpret this
\nct{es04}
through the coherent combination of radiation in two natural propagation 
modes, with some relative phase delay. 
Melrose et al.~(2006), however, attribute this effect to incoherent
\nct{mmk2006}
combination of specific distributions of radiation in both modes. 
Both these so different models focus on the single pulse PA behaviour 
at a fixed pulse longitude, and achieve reasonable consistency with data.

Contrary to that approach, in this paper I interpret variations 
of the non-RVM effects across the pulse window, i.e.~the goal is to 
understand the distorted shape of complex PA curves. The statistical single pulse 
properties need to be taken into account 
in such analysis, but fortunately this can be done through their simple 
parametrisation.
Sect.~\ref{obser} describes observations that can be considered a clear
manifestation of the coherent addition of modes, 
and which inspired the analysis of this paper. 
After introducing the needed mathematics (Sect.~\ref{model}), 
the coherent mode
addition is applied to a few peculiar polarisation effects
(Sect.~\ref{results}).

\section{Observational evidence for the coherent addition of modes}
\label{obser}

A strong signature of the coherent mode addition (CMA), 
far more convincing than the mere 
existence of $V$, is presented by those 
orthogonal PA transitions,
that are accompanied by high levels of $|V|$. Especially those 
which occur at the peak of $|V|$. 
This type of
behaviour suggests CMA, because it is expected in the case of 
orthogonal modes added with a phase lag which changes
with pulse longitude.

\begin{figure}
\includegraphics[width=0.48\textwidth]{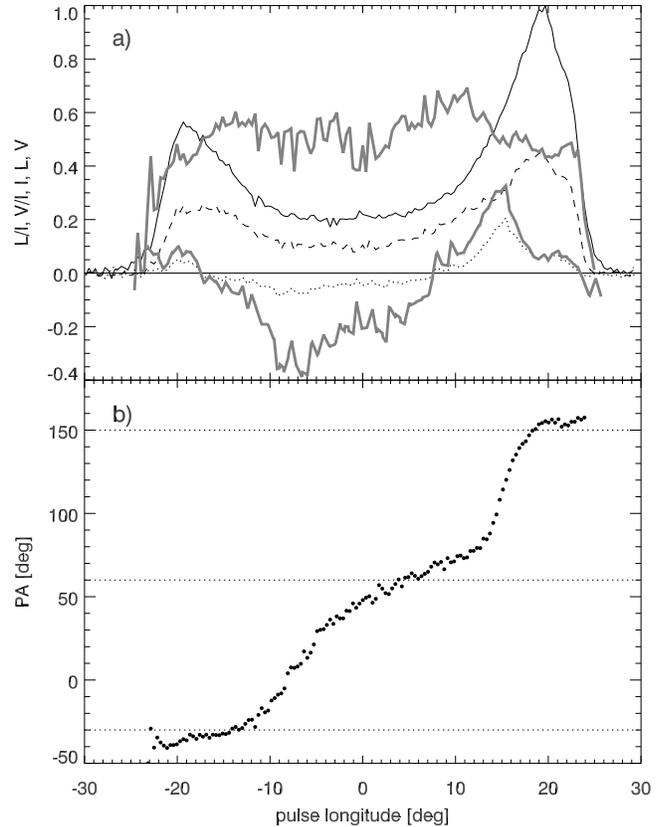}
\caption{Polarised profile of PSR B1913$+$16 (after Weisberg \& Taylor 2002). 
The polarisation fractions $L/I$ and $V/I$ are shown with thick grey lines. 
Note the three flattenings of PA separated vertically 
by about $\sim90^\circ$. 
The PA transitions at $\Phi\approx -7^\circ$ and $15^\circ$ coincide with
maxima in $|V|/I$, which contradicts the traditional 
(incoherent addition) interpretation. 
}
\label{weisberg}
\end{figure}

A clear example of such phenomenon is provided by the profile of 
the relativistic binary pulsar B1913$+$16 (Fig.~\ref{weisberg}, based on
Fig.~7 in  
Weisberg \& Taylor 2002). \nct{wt02}
The PA of this pulsar exhibits a continuous increase across a larger than $\pi$ interval of 
195$^\circ$.
On the inner side of two maxima that flank/surround the profile, the PA
makes two steep transitions between three approximately
orthogonal values. Both these PA transitions
are accompanied by a high
level of circular polarisation ($|V|/I \sim 0.3$) 
which peaks at pulse longitudes 
where the PA assumes a midpoint value, located half way between 
the nearly constant values of apparently pure orthogonal modes.

Several properties of these PA transitions differ strikingly from 
the well known properties of transitions that are produced by 
the incoherently combined modes. The latter appear when the amount of one
mode starts to exceed another one, and are expected to occur at longitudes 
where $L/I$ drops to zero and $V$ crosses zero on its way to change
sign.
In B1913$+$16 the linear polarisation fraction (top thick grey line) 
does not even
reach the proximity of zero. Neither does the circular 
polarisation change sign in coincidence with the PA transition. 

Another example of similar effect is provided by PSR B1933$+$16.
As can be seen in Fig.~1 of MRA16 
\nct{mra2016}
at the longitude of 
$-5^\circ$ the PA curve is almost transiting to another polarisation mode.
Linear polarisation degree is fairly low at the jump, however, $|V|$ 
again reaches a maximum on the way between the orthogonal modes 
(instead of crossing zero). 
The average PA does not quite reach 
the other mode, and immediately retreats (at longitude of $-4^\circ$), 
to form 
a narrow V-shaped distortion of the PA curve. The feature makes
impression of a failed mode jump, or some superficially similar jump-like 
behaviour.

Even more striking view of this jump, however, is provided by the dotty 
distribution of PA recorded in individual samples. The PA 
distortion is possibly bidirectional, 
with the PA of some samples increasing,
while most of them have the PA decreasing in accordance with the average PA
curve. As a result, the V-shaped distortion of the average PA 
forms only a lower half of a loop-like structure, 
created by the PA of
individual samples. 

Just four degrees ahead of this loop-shaped distortion in B1933$+$16,  
the profile presents a well behaving, regular orthogonal mode jump, with
little $V$ or $L$, i.e.~consistent with the ``incoherent" mode summation 
(coherent sum of orthogonally-polarised waves of equal amplitude 
generally does not make $L$ vanishing). Note that `incoherent' means coherent 
addition of waves with a large mixture of different phase lags, sampled from
a wide phase lag distribution (wider that $2\pi$).
If the loop-like (or V-shaped) distortion is interpreted as a 
coherent sum of modal contributions, the observed signal would have to 
quickly change
(within $4^\circ$ of pulse longitude) from an incoherent sum of modes 
to a coherent sum of 
modes. This suggests it may be worth to attempt to interpet the whole 
pulsar emission in terms of CMA, albeit with longitude-dependent width of 
the phase lag distribution. 

Further examples of similar phenomena include the infamous PA distortions
observed within core components of M-type pulsars: B1237$+$25 (SRM13) 
and B1857$-$25 (Mitra \& Rankin 2008). 
\nct{srm13, mr2008}
With its loop-like shape, 
the distortion in B1237$+$25 is superficially similar 
to that in B1933$+$16. However, there are two key differences: 
1) In B1237$+$25 $V$ has an antisymmetric sinusoid shape, with a sign
change near the middle of the core component (in B1933$+$16 $V$ is negative 
throughout the loop). 2) In B1237$+$25 the PA loop is created 
as a split of the primary mode PA track, however, the loop converges 
back to the secondary mode, which is visible in the peripheries of the profile 
in the form of short patches of an orthogonal PA. In B1933$+$16 
the PA loop opens and closes at the  primary mode.
Remarkably, in these complicated-core objects, $V$ changes sign not at 
the minima of $L$. The minima are located on both sides of the disappearing
$V$, and again seem to be roughly coincident with peaks of $|V|$. 

\section{The model}
\label{model}

\subsection{Simple model of coherent mode addition}

I assume that radiation observed at any pulse longitude 
results from coherent 
combination of linearly polarised natural mode waves. These are
represented as two monochromatic sinusoids:
\begin{equation}
E_x = E_1\cos{(\omega t)}, \ \ \ E_y = E_2\cos{(\omega t -
\Delta \phi)},
\label{waves}
\end{equation}
where $\Delta\phi$ is the phase delay between the modes. 
The Stokes parameters for such waves 
are calculated from:
\begin{eqnarray}
I  & = & E_1^2 + E_2^2\\
Q & = & E_1^2 - E_2^2\label{qst}\\
U & = & 2 E_1 E_2 \cos(\Delta\phi)\label{ust}\\
V & = & -2 E_1 E_2 \sin(\Delta\phi)
\label{stokes}
\end{eqnarray}
(Rybicki \& Lightman 1979). The  \nct{rl79} 
polarisation fractions and
angle are calculated in the usual way:
\begin{eqnarray}
\Pi_{\rm tot}  =  I_{\rm pol}/I & = & (Q^2 + U^2 + V^2)^{1/2}/I\\
\Pi  =  L/I &  =  & (Q^2 + U^2)^{1/2}/I\\
 \psi& = & 0.5\arctan{(U/Q)}.\label{psiang}
\label{psaj}
\end{eqnarray}

It is assumed that 
the observed polarisation results from an intrinsic signal 
(likely produced directly by the emission mechanism)
which has propagated through some intervening magnetospheric plasma, 
possibly located at the polarisation limiting radius 
(PLR, e.g.~Petrova \& Lyubarskii 2000). \nct{pl00}
The intervening matter is represented by a polarisation basis
$(\vec x,\vec y)$ and the signal enters the basis at a
polarisation angle $\psin$ given by:
\begin{equation}
\tan{\psin}=E_2/E_1=R,
\end{equation}
where $R$ is the mode amplitude ratio,  
\begin{equation}
E_1=E\cos\psin{\rm,\ \ \ \ \ \ }E_2=E\sin\psin,
\label{amplits}
\end{equation}
and $E$ is the incident wave amplitude.
The PA observed in single pulses typically exhibits wide distributions
at a fixed longitude, which are assumed to be caused by the 
stochastic nature of the input signal. I assume that typically a wide
distribution of $\psin$, denoted $\npsi$, 
which at some longitudes may possibly be even quasi-isotropic, 
enters the polarisation basis (different $\psin$ contribute 
at a given pulse longitude in different single
pulses). The PA at which $\npsi$ is maximum is denoted by
$\ppk$, and the width of $\npsi$ by $\wpsi$ . 

The incident signal 
is split into components $E_x$ and $E_y$ parallel 
to the polarisation direction of the proper propagation modes 
in the intervening matter, which induces some phase lag $\Delta\phi$ 
between $E_x$ and $E_y$. 
The intervening matter 
is assumed to be highly modulated and 
irregular, hence the lag is also expected to be highly random, and
represented by a wide distribution $\nlag$ with the maximum at
$\lpk$ and the width $\wlag$. 
The whole propagation physics is parametrised with these two quantities.

\begin{figure*}
\includegraphics[width=0.78\textwidth]{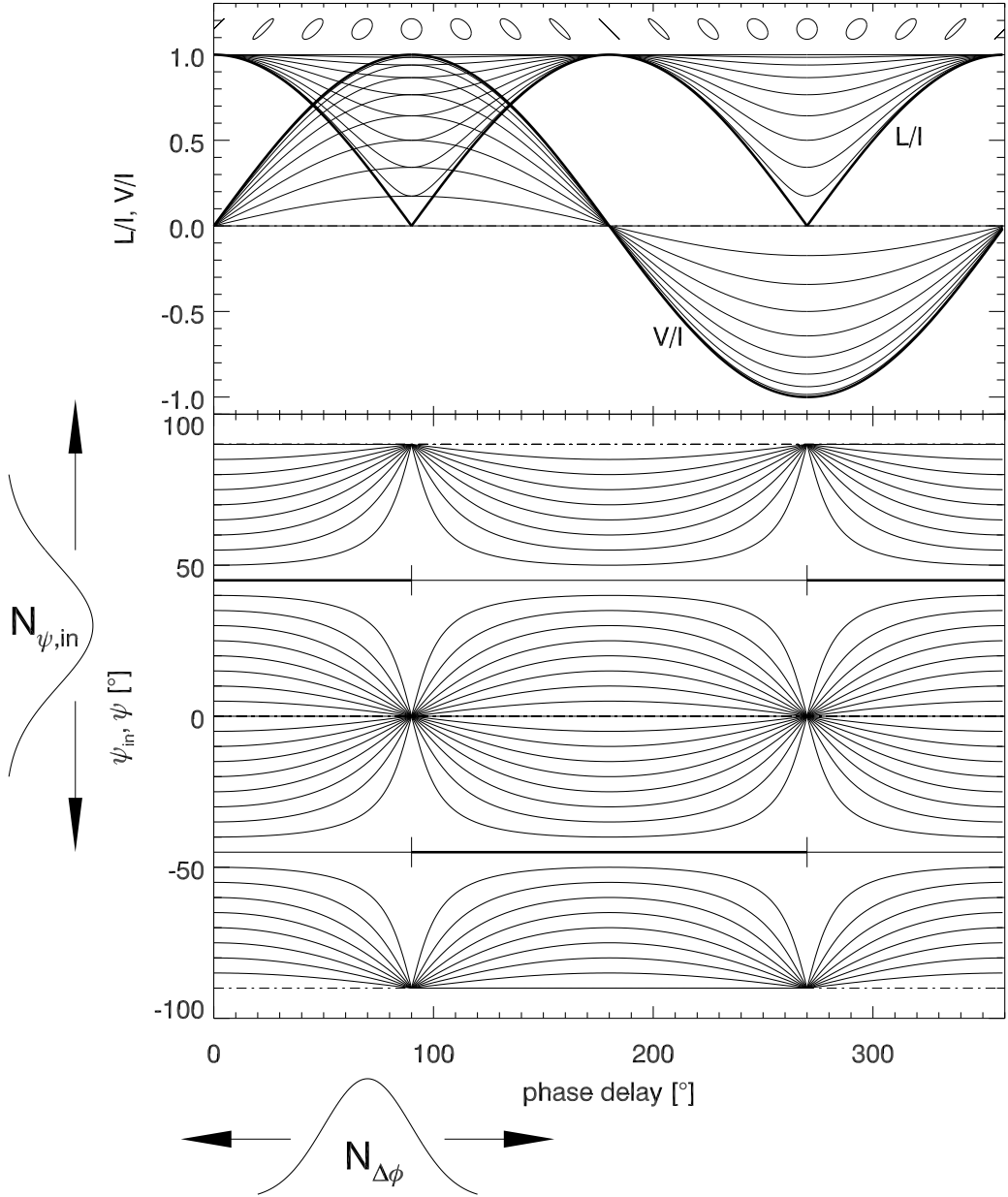}
\caption{Polarised fractions $L/I$ and $V/I$ (top) and polarisation angle
$\psi$ (bottom) that result from coherent addition of two 
orthogonally polarised waves with different amplitude ratio $R$ and phase lag 
$\lag$. 
$R$ is fixed along each line in the bottom panel and equal to $\tan\psin$,
with the values of the incident angle $\psin=\psi(\lag=0)$ separated by
$5^\circ$. The observed OPMs are visible as the thick black spots
at which the
lines of fixed $\psin$ converge. Statistical distribution $\npsi$ 
is shown on the left, and $\nlag$ -- below the bottom panel. 
$L/I$ and $V/I$ are presented only for $\psin\in[0,45^\circ]$. For
the increasing $\psin$, $L/I$ evolves from a constant ($L/I=1$ for $\psin=0$) 
to the form of $|\cos\lag|$ for $\psin=45^\circ$, 
whereas $V/I$ -- from a constant ($V/I=0$) 
to the $\sin\lag$ shape. Thick lines in both panels and the polarisation
ellipses on top present the case of 
equal mode amplitudes ($\psin=45^\circ$). Note the anticorrelation of $L/I$
and $|V|/I$.
Variations of the polarised fractions within the
full range of $\psin$ are shown in bottom panels of Fig.~\ref{snake}.
}
\label{modes}
\end{figure*}

It is assumed that the observed variations of polarisation as a function of
pulse longitude\footnote{The capital letter $\Phi$ refers 
to the pulse longitude 
in the profile,
and should not be mistaken with the small $\phi$ which denotes oscillation 
phase in the wave given by eq.~(\ref{waves}). 
Hereafter I always use the word `longitude' 
to refer to $\Phi$,  whereas `phase' is used for $\phi$.
}
$\Phi$, mostly result from the longitude dependence 
of the mode ratio and phase lag, i.e.:
\begin{eqnarray}
E_2/E_1&=&\tan\left(\ppk(\Phi)\right)
\label{pierwsze}\\ 
\wpsi&=&\wpsi(\Phi)\\
\lpk&=&\lpk(\Phi)\\
\wlag&=&\wlag(\Phi).
\label{ostatnie}
\end{eqnarray}
The function $\ppk(\Phi)$ is determined by the relative 
orientation of a distribution of incident PAs and the intervening basis, 
as described in Section \ref{picture}. Since the emission and propagation 
physics is neglected, the other functions need to be found by a guess, 
hopefully guided by the qualitative properties of available data. 
Equations (\ref{pierwsze})-(\ref{ostatnie}) make
the Stokes parameters dependent on the pulse
longitude.

Hereafter I will be interested in polarisation
fractions and angle, since they depend only on the ratio $R=E_2/E_1$. 
This makes it possible to ignore the absolute profiles of 
$I$, $L$ and $V$, i.e.~the incident wave amplitude $E=1$
in eqs.~(\ref{amplits}).\footnote{Unfortunately, in observational 
works 
usually only the observed $I$, $L$ and $V$ are published.  
This makes it very difficult to 
assess $L/I$ and $V/I$ whenever $I$ is changing steeply or is very low.  
This plotting tradition
seriously hampers any attempts to interpret the observed pulsar polarisation.}

The polarisation angle $\psi$,  as given by
eq.~(\ref{psaj}), so far only includes the effects of mode addition, without 
the likely variations caused by the changes of projected magnetic field
 with $\Phi$. If these regular RVM variations of PA
are denoted as $\psi_{\sss RVM}(\Phi)$, the observed PA is:
\begin{equation}
\psi_{\rm obs} = \psi + \psi_{\sss RVM},
\label{totpsaj}
\end{equation}
i.e.~$\psi_{\sss RVM}$ serves as a reference value for 
the non-RVM effects.

\subsection{The origin of orthogonal polarisation modes}


It is then assumed that the input wave becomes 
 split into components parallel to the natural polarisation directions, 
then the components acquire some phase lag $\lag$, and finally combine. 
Fig.~\ref{modes} presents the polarisation characteristics for 
 $\lag$ values within one full wave oscillation period, and for a set  
 of  mode amplitude ratios that correspond to the input PA values separated by
$5^\circ$. The bottom panel presents the PA as a function of the 
phase lag, with $\psin$ fixed on each curve of $\psi(\lag)$.
 For each point in the bottom panel, $\psin$ 
can be established by following the closest curve of $\psi(\Delta\phi)$ 
towards the left vertical axis,
i.e.~$\psi_{in}= \psi(\Delta\phi=0)$. 
The PA of the natural modes of the intervening polarisation 
basis is shown with the dot-dashed horizontal lines 
at $\psi=\pm90^\circ$ and $0^\circ$ (the latter is overlapping with
$\psi(\psin=0)$). The straight lines that correspond
 to equal mode amplitudes
 (incident PA of $\pm45^\circ$, hereafter called intermodes) 
consist of separate sections that jump discontinously 
to orthogonal values at $\lag=90^\circ$ and
$270^\circ$ (as marked with ticks and thick line sections).
This corresponds to the transformation of the polarisation ellipse 
into a circle, and further into an orthogonally oriented ellipse 
(as shown near the top of the figure). 
Changes of $L/I$ and $V/I$ for this equal amplitude case ($\psin=45^\circ$) 
are shown in the top panel with thick lines, which clearly demonstrate 
the anticorrelation of $L/I$ and $|V|/I$ inherent for the CMA.


The orthogonal polarisation modes of radio pulsar emission 
clearly stand out at the locations 
where several different curves of $\psi(\Delta\phi)$ 
cross at the same point. In single pulse observations, 
at a fixed pulse longitude,  
different points of this figure are quasi randomly sampled, with some spread 
in the input PA ($\npsi$, shown left of the plotting box) 
and some spread in the
phase lag ($\nlag$, shown below the plot).

The origin of these OMPs, emerging as the nodes 
in the lag-amplitude ratio diagram, is simple: the input signal which is 
polarised at an arbitrary $\psi_{in}$, produces polarisation ellipses 
parallel to either natural polarisation mode, whenever the lag 
$\Delta\phi=\pi/2$. 
That mode will be observed, which corresponds to the larger component 
of the split input vector.

\begin{figure}
\includegraphics[width=0.48\textwidth]{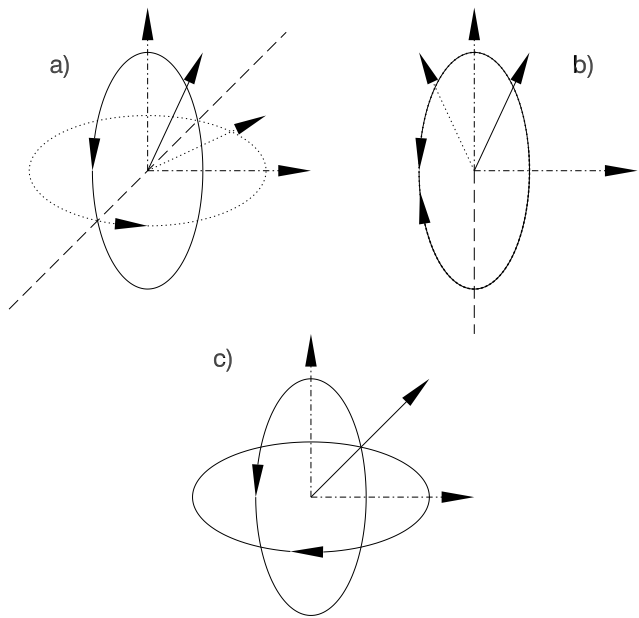}
\caption{
Top: polarisation ellipses for coherent addition of orthogonally-polarised waves 
with the relative phase lag of $\lag=\pi/2$. The waves' amplitudes 
correspond to projections of either the solid or the dotted incident vector 
on the dot-dashed basis vectors. 
 Dashed line marks the orientation of a peak of polarisation vector 
distribution $\npsi$, from which the two vectors are drawn.  
Arbitrary orientation of the input vector in {\bf a} always gives the PA 
of either orthogonal mode.
The solid
vector in {\bf a} produces the solid polarisation ellipse of the same
handedness as the dotted ellipse of the dotted vector, hence $V$ peaks 
when $\npsi$ is inclined at $45^\circ$ with respect to the basis.
When one mode dominates i.e.~$\npsi$ is
aligned with one basis vector as in {\bf b}, the handedness of ellipses is
opposite and $V$ vanishes. Bottom: for equal amount of incoherently 
summed elliptical OPMs ({\bf c}) the handedness is opposite (unlike in {\bf
a}) and $V$ vanishes, whereas $V\ne0$ when one elliptical mode dominates.
}
\label{ellipses}
\end{figure}

\begin{figure*}
\includegraphics[width=0.68\textwidth]{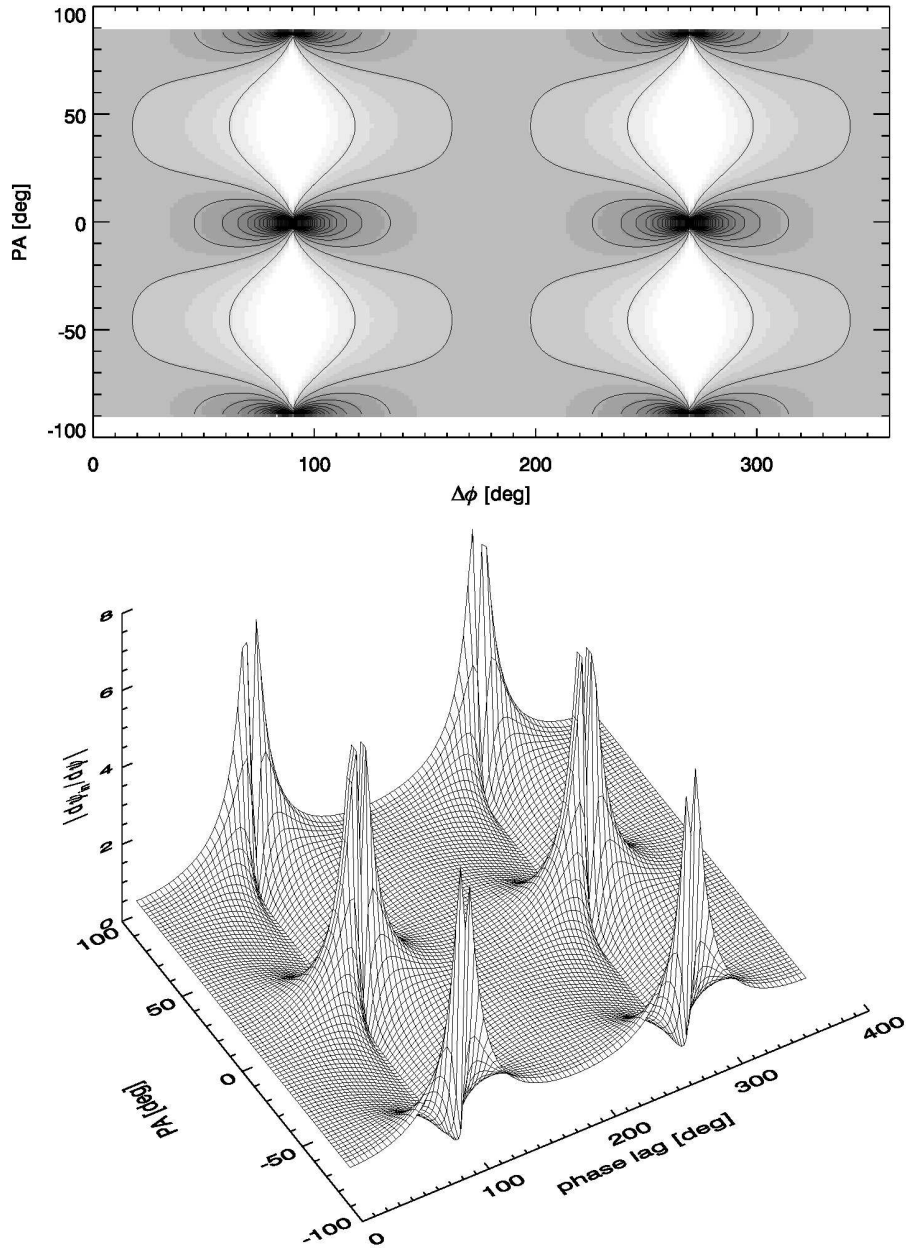}
\caption{Probability $P$ of measuring the polarisation angle $\psi$ 
for the random $\npsi$ and $\nlag$. The OPMs are visible as the black spots
(top) or the sharp projections (bottom). The height of the spikes is
arbitrarily determined by the numerical grid density.
}
\label{cienie}
\end{figure*}

The phenomenon is presented in Fig.~\ref{ellipses}a, in which components of 
a dotted electric field vector,
tilted at an arbitrary input angle of $\psi_{in}=25^\circ$ 
combine at $\Delta\phi=90^\circ$ 
into the dotted polarisation ellipse, 
whereas another input vector at $\psi_{in}=65^\circ$ (solid), similarly
produces the solid ellipse. This way the two orthogonal modes, 
always polarised
at $0^\circ$ or $90^\circ$ are produced.\footnote{The PA,
$L$, and
$V$ of the elliptically polarised radiation  are determined assuming 
decomposition into 
the fully circular part, and the phase coherent linear part in the direction of
the major axis of the ellipse.} 
If $\lag$ is slightly different from $90^\circ$, the polarisation
ellipses are no longer strictly parallel to the $\vec x$ or $\vec y$ 
vector of the basis, however, for most $R$ the misalignment is small 
(this can be seen 
by making a vertical cut through Fig.~\ref{modes}, e.g.~at $\lag=80^\circ$). 
If the input vectors appear on both sides of the dashed diagonal in 
Fig.~\ref{ellipses}a, 
the two orthogonal modes will be present in the observed signal
whenever the medium imposes phase delays 
in the vicinity of $\pi/2$ on the propagating
 signal. If $\Delta\phi\ne \pi/2$, variety of PA values will be observed, 
however, their fixed-longitude distributions will reveal maxima at the modes.
This can be understood by considering the local line density along some
vertical cut of Fig.~\ref{modes} 
(any vertical slice which is not at a multiple of
$\Delta\phi=\pi$). Thus, the modes will emerge statistically in
the observations, even for phase lags much different from $90^\circ$, 
although in such case they will form a distribution with wide peaks 
connected with bridges.
Therefore, if the lag and $\psin$ are sampled from a wide or random
distribution, the modes stand out clearly in statistical sense, i.e.~appear
frequently.

When both the distributions 
($\npsi$ and $\nlag$) are uniform, 
the bottom panel of  Fig.~\ref{modes} will not be uniformly covered by
samples. Instead, the PA observed in different samples 
 will bunch at the orthogonal modes, following the density 
of the lines in the figure. In spite of the pure randomness 
of input polarisation parameters, 
the orthogonal modes will anyway 
stand out in single pulse observations with equal strength, but the average
profile will be fully depolarised.
The apparent OPMs generated through the coherent mode combination 
are then an extremely 
robust feature, easily surviving most of sampling styles that can be
imagined for Fig.~\ref{modes}. Only the 
zero lag, or
a multiple of $\pi$, does not produce the OPMs. 

In other words, the density of lines in Fig.~\ref{modes} presents
the probability of measuring a given PA for a totally random 
orientation of the input polarisation and for a random 
(uniform) distribution of phase lags (by `measuring' I mean 
the measuring of PA at short time samples always taken at a fixed pulse
longitude).  Equations \mref{psiang}, \mref{qst}, and \mref{ust} give
\begin{equation}
K\equiv\frac{1/R-R}{2}=\frac{\cos\lag}{\tan2\psi}.
\end{equation}
A square equation $R^2 + 2KR - 1 = 0$ gives the solution for 
$\psin(K)$, 
which provides the input PA for any point in Fig.~\ref{modes}:
 $\psin(\lag, \psi)$.  
The above-described probability distribution 
(for the uniform $\npsi$ and $\nlag$)
is then calculated as the rate at which the lines of fixed $\psin$
are crossed vertically: 
\begin{equation}
P(\psi,\lag)_{rndm} = \left|\frac{d\psin}{d\psi(\lag)}\right|
\end{equation}
and is presented in Fig.~\ref{cienie}.

If 
$\nlag$ in Fig.~\ref{modes} is narrow, whereas 
$\npsi$ is very wide (or isotropic) then  
clearly defined OPMs will be observed
when $\nlag$ 
peaks near $\lag=90^\circ$ or $270^\circ$. 
If the peak of $\nlag$ is at $\lag=0^\circ$,
$180^\circ$, or a multiple of thereof, the quasi-uniform distribution of
the input PA will be observed. 
Intermediate cases produce wide observed PA
distributions, with broad tracks of quasi-orthogonal modes, 
and many PA values between
them. Inclusion of a narrow PA distribution (shown on the left in
Fig.~\ref{modes}) 
further modifies the picture, as will be described below.

For a symmetrical orientation of input vectors 
with respect to $\psi_{in}=45^\circ$, equal amounts of same-handedness 
circularly polarised signal are produced (Fig.~\ref{ellipses}a).
It is important to note that this case is different from the case 
when the input signal is split into two natural orthogonal modes which
are elliptically polarised and reach the observer without 
being combined on the way. In the last case the handedness of $V$ is
opposite (Fig.~\ref{ellipses}c) and the resulting $V=0$. 
A symmetrical distribution of the input PA, 
centered at $\psi_{in}=45^\circ$
(marked with the dashed line in Fig.~\ref{ellipses}a) will produce equal 
amounts of the two OMPs plus strong $V$. The same distribution 
centered at $\psi_{in}=90^\circ$ (or $0^\circ$) will produce one strong 
polarisation mode with little $V$, as illustrated in Fig.~\ref{ellipses}b. 
This is again different from the elliptical mode case, in which a single mode
signal should reveal the elliptical polarisation intrinsic to the natural 
propagation mode.

It is a notable feature that for the vectors located on different sides 
of $\psi_{in}=45^\circ$ (Fig.~\ref{ellipses}a) the orthogonal OPMs 
are produced \emph{with the same handedness of circular polarisation}.
This phenomenon is observed in the single pulse emission, 
where samples of the same handedness appear on both orthogonal mode PA
tracks. This is well illustrated in Figs.~1 and 2 of Mitra et al.~(2015) 
who plot opposite-handedness samples 
in different colors. 

Another very important point is that in the case 
of the coherent mode addition
 $V$ changes sign when the peak of the input PA distribution 
moves across one of the natural modes (Fig.~\ref{ellipses}b). Therefore, 
different samples of the same polarisation mode may have 
opposite sign of $V$, as can also be seen in Figs.~1 and 2 of MAR15.

Both the above-described properties are opposite to what is known about the 
noninteracting elliptically polarised propagation modes.

As shown in Fig.~\ref{ellipses}a the answer to the question of 
which mode will be detected, depends on the orientation of the input
polarisation vector with respect to the separating angle of $45^\circ$. 
Accordingly, in the case of symmetrical $\npsi$ distributions that peak at
$\ppk$, 
the predomination of a
given mode depends on the position of $\ppk$ relative to $45^\circ$. 

To learn what Stokes parameters contribute at a given pulse longitude,
$\npsi$ and $\nlag$, shown on the margins of Fig.~\ref{modes}, must be 
convolved with the line density distribution on the lag-PA diagram.  
The convolution of $\nlag$ 
is simple: the 
lag distribution selects the whole vertical area in Fig.~\ref{modes},
located directly above $\nlag$, and the contribution of 
different phase lags to the average signal is weighted by $\nlag$.
The PA distribution on the left, however, is convolved 
indirectly: for the input PA of, say, $40^\circ$, it is
necessary to start at this value on the left axis of Fig.~\ref{modes},
and follow the fixed $\psin$ line rightwards, 
until the vertical region selected by the lag
distribution is reached. This implies that whenever the PA distribution 
extends to both sides of $\psin=45^\circ$, 
both OPMs are produced and they may appear in single pulse
data (at a given phase in different rotation periods, 
i.e.~they may be 
visible as the parallel tracks of orthogonal PA).\footnote{The visibility 
of both OPMs  
is known to depend on the question of whether the two OPMs, i.e.~the 
different results of the coherent addition, 
can simultaneously contribute to a single time sample (Stinebring et
al.~1984; McKinnon 
\& Stinebring 2000).
\nct{scr84, ms00}}
For $\npsi$ which
does not extend beyond  the interval of $[-45^\circ,45^\circ]$ 
(or beyond $[135^\circ, 215^\circ]$), a single OPM will be
observed. 

\begin{figure}
\includegraphics[width=0.48\textwidth]{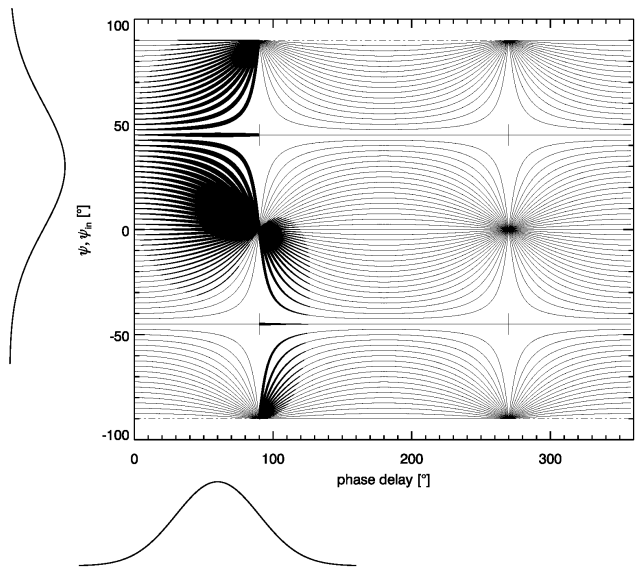}
\caption{Effect of convolution of moderately narrow Gaussian distributions 
$\npsi$ 
and $\nlag$ on the lag-PA diagram. Thickness of the fixed-$\psin$ lines 
is proportional to the product $\npsi(\psin)\nlag(\lag)$. Both
quasi-orthogonal polarisation modes 
appear (near $\psi=0$ and $\pm90^\circ$) because the wing of $\npsi$, centered at $\ppk=30^\circ$, 
extends across $\psin=45^\circ$. The other parameters are: 
$\sigma_{\psi, in}=30^\circ$, $\lpk=60^\circ$, $\sigma_{\lag}=30^\circ$.
}
\label{ortho3}
\end{figure}

An arbitrary example of combined effects of $\npsi$ and  $\nlag$ 
is shown in Fig.~\ref{ortho3}, where a moderately wide 
PA distribution peaking at 
$\psin=30^\circ$, and a lag distribution peaking at 
$\lag=60^\circ$, generate two approximately orthogonal
polarisation modes of unequal magnitude (at $\Delta\phi\approx90^\circ$). 
The thickness 
of $\psi(\lag)$ curves in that figure is made proportional to 
 the product of appropriate values in the distributions: 
$\npsi(\psin)\nlag(\lag)$, where $\psin$ 
should be understood as $\psin=f(\psi,
\lag)$. 
The imperfect orthogonality of OPMs, often observed in pulsars,  
is a natural consequence of the convolution presented in Fig.~\ref{ortho3}.

The PA and lag distributions do not necessarily 
need to be as narrow as shown
in Fig.~\ref{modes}. 
In particular, $\nlag$
can span much more than $2\pi$. The resulting
polarisation characteristics are then a residual effect 
determined by the horizontal misalignment of $\nlag$ with respect
to vertical symmetry lines of the lag-PA pattern 
(e.g.~$\Delta\phi=90^\circ$, or
$180^\circ$). Possible asymmetry of the lag distribution wings could also be
in such case decisive for the ensuing PA.

\section{Results}
\label{results}

The OPM model of previous section will now be employed to 
understand selected examples of
pulsar polarisation zoo. 

\subsection{The origin of polarisation in B1913$+$16}

\begin{figure*}
\includegraphics[width=0.88\textwidth]{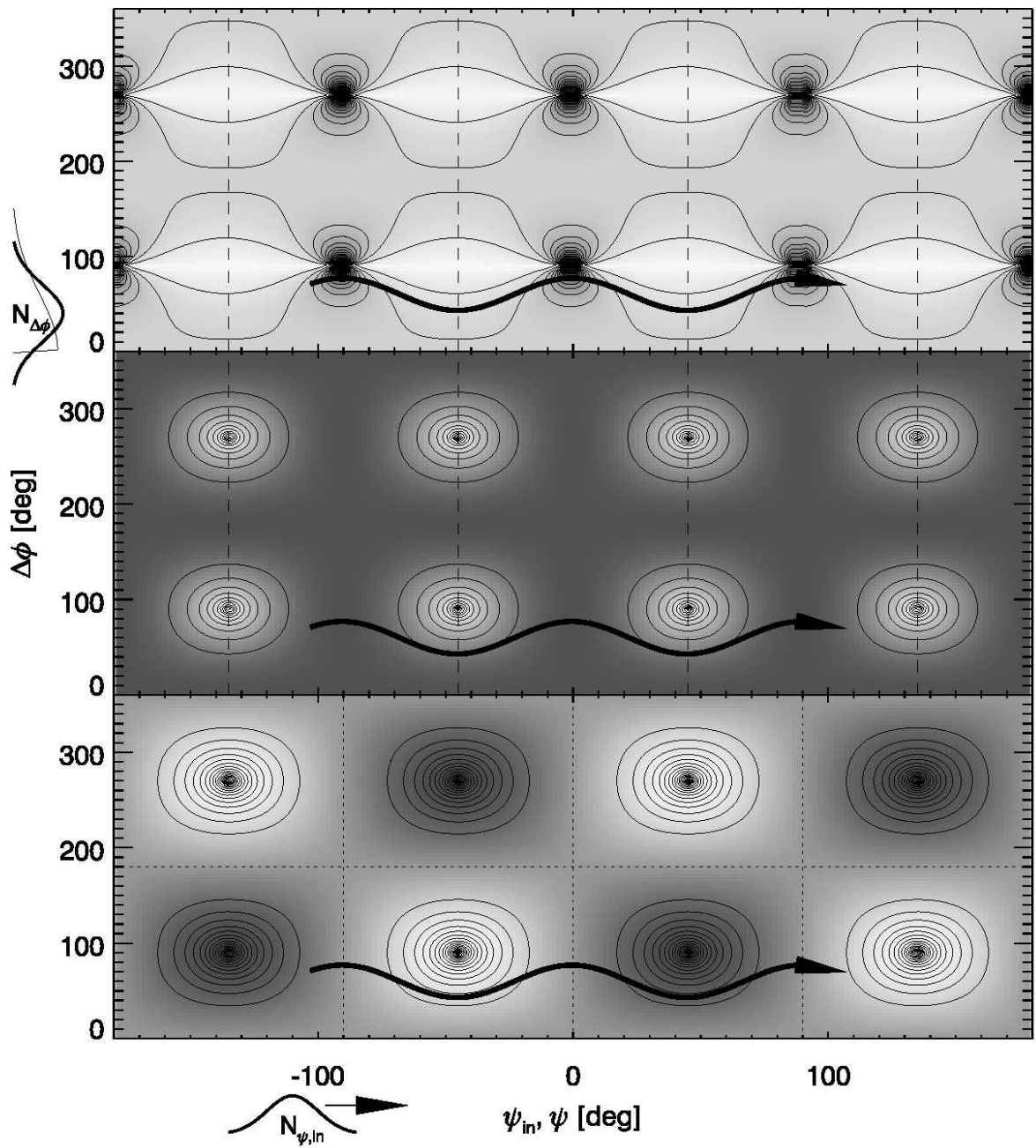}
\caption{A diagram useful for interpretation of pulsar polarisation,
especially in cases when $\npsi$ and $\nlag$ are narrow. 
Top: the probability $P(\psin,\lag)$ of measuring the PA marked on the horizontal axis, 
assuming uniform distributions of $\npsi$ and $\nlag$. 
Note the different layout, with $\psin$ on the horizontal axis, and $\lag$ on
the vertical axis.
Middle: 
linear polarisation fraction $\Pi(\psin, \lag)$. Bottom: circular
polarisation fraction $V/I$ as a function of $(\psin, \lag)$. The value of 
$P$, $\Pi$ and $V/I$ increases in darker regions (bright $V/I$ regions are
negative). Dashed verticals mark the intermode PA ($45^\circ$ away 
from the modes at $0$, and $\pm90^\circ$). Dotted lines separate $V$ of
opposite sign. For more details see text.
}
\label{snake}
\end{figure*}

The PA curve of B1913$+$16 (Weisberg and Taylor 2002, reproduced here in
Fig.~\ref{weisberg}), if considered alone, looks quite innocuous: 
there are three PA flattenings that
seem to present the two natural OPMs separated by two gradual
transitions between them. 
However, $|V|/I$ astonishingly peaks 
at the transitions 
(instead of crossing zero), 
and both the transitions are gradual. Equally strange, $L/I$ is 
high everywhere (top grey line in Fig.~\ref{weisberg}). 
Instead of reaching zero, $L/I$ decreases near the trailing PA jump from
$0.65$ to $0.45$.

Since generally $|V|/I>0$ in the profile, the peak $\lpk$ of the $\nlag$
distribution must either be displaced from zero, or the distribution must be
asymmetric, because otherwise $V$ would vanish (e.g.~see Fig.~\ref{ellipses}b, 
and note that a change of the lag from $\pi/2$ to $-\pi/2$ is equivalent 
to a change of sign in one component 
of the incident vector).
Two examples of such distributions (denoted $\nlag$) 
are shown on the left margin of Fig.~\ref{snake}.
The asymmetric $\nlag$ with only positive $\lag$ 
is expected for the PLR scenario (Cheng \& Ruderman 1979; Lyubarskii \&
Petrova 1998; ES04) 
\nct{cr79, lp98, es04}
although both the distributions often 
produce similar results if the symmetric one is not too wide and is  
displaced to some 
$\lpk>0$.\footnote{The observation of slightly 
elliptical modes by ES04 in B0329$+$54 is not necessarily 
 inconsistent with the PLR effects on the purely linear natural modes, 
because  the observed 
OPMs are 
the net result 
of the coherent combination of the natural propagation modes with some 
$\npsi$ and $\nlag$. A slight asymmetry  of $\npsi$ (and $\nlag$) around zero 
is sufficient for 
the observed mode to acquire some ellipticity even when the natural modes 
are perfectly linear (Fig.~\ref{ellipses}b). 
Still, the natural 
elliptical modes are supported by the observed 
OPM jumps at which $V$ does change the sign. 
For simplicity I ignore this ellipticity
in this paper.
}

If the small positive $V/I$ on the leading edge of the profile is ignored 
($\Phi\approx-20^\circ$ in Fig.~\ref{weisberg}),
a sinusoid-like $V/I$ is observed in the rest of the pulse window, i.e.~$V/I$ 
decreases from zero to a negative minimum at $\Phi=-7^\circ$, crosses zero 
at $7^\circ$, reaches maximum at $\Phi=15^\circ$ and finally drops to zero at the
trailing edge.
Since $V$ should change sign whenever $\ppk$ coincides with one of the natural 
modes, the observed $V$ profile suggests that 
the input vector must rotate by slightly more than $180^\circ$, say between 
the values of $-100^\circ$ and $100^\circ$ in Fig.~\ref{snake}.
This interval of $\psin$ is hereafter assumed to be cast
onto the observed pulse window of B1913$+$16.
The initial position of the $\npsi$ distribution 
is shown at the bottom of the figure, and 
it is assumed to roughly correspond to the leading edge of the observed
profile. While the pulsar rotates, 
$\npsi$ moves rightwards. 

The top panel of Fig.~\ref{snake} shows the probability distribution that needs
to be convolved with $\npsi$ and $\nlag$ to 
determine the polarisation characteristics at a given longitude.
In a general case of arbitrary distributions, 
the procedure is as follows: for each pulse longitude 
(i.e.~for a given position of
$\npsi$ at $\ppk$) the Stokes parameters, as determined by $\lag$ and
$\psin(\lag,\psi)$, need to be integrated over 
$\nlag$ to obtain the observed distribution of PA (i.e.~$\psi$) at a
fixed $\Phi$ (or distributions of $L/I$ and $V/I$ at that $\Phi$). 
To obtain a single value of PA in an average profile, the integration needs
to be done over both $\nlag$ and $\npsi$.

For B1913$+$16 I employ moderately narrow distributions, which makes it 
possible to read out the result directly from Fig.~\ref{snake}.
The modal peaks, located at $\psin=0$ and $\pm90^\circ$ in the $P$ distribution
(top panel) are very strong features. Therefore, when $\npsi$ (i.e.~$\ppk$) 
coincides 
with one 
of the modal PA values at some pulse longitude $\Phi$, 
that OPM value statistically 
dominates in single pulse samples observed at that $\Phi$. In other words, 
the peak of the convolved net probability distribution 
will be located close to the modal peaks. 
However, if $\npsi$ is located at an intermode (e.g.~$\ppk=45^\circ$),
then the modal points are located in the low-level 
wings of both $\npsi$ and $\nlag$.
 The strongest (most likely) contribution then comes from the location in the 
$(\psi, \lag)$ diagram which corresponds to the peaks of $\npsi$ and
$\nlag$. For the symmetric $\nlag$ distribution shown on the left margin of
 Fig.~\ref{snake}, this corresponds roughly to
$(\psi,\lag)=(45^\circ, 40^\circ)$. 

\begin{figure}
\includegraphics[width=0.48\textwidth]{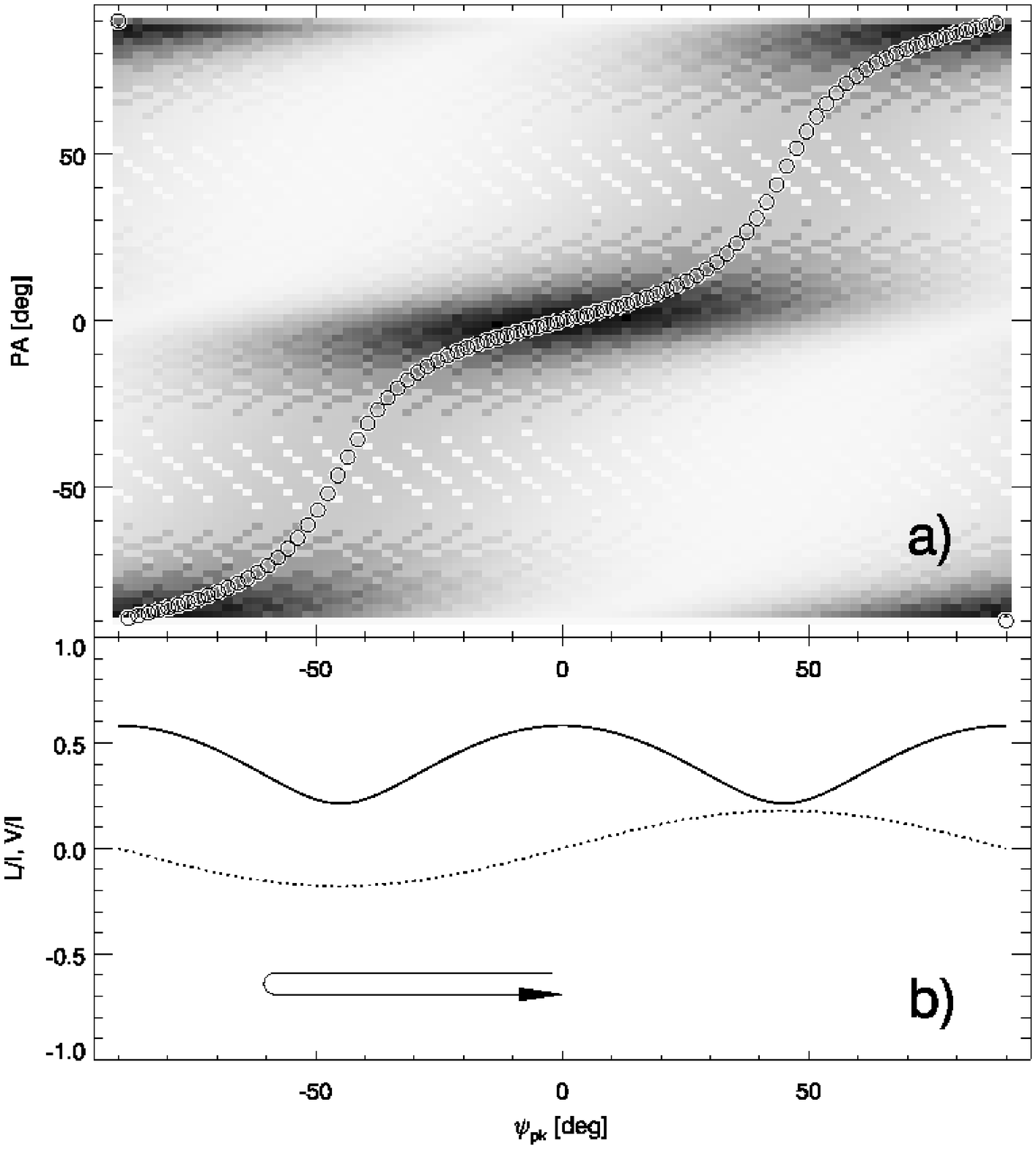}
\caption{Polarisation characteristics modelled for PSR B1913$+$16 
(cf.~Fig.~\ref{weisberg}).
The span of $\ppk$ on the horizontal axis 
corresponds to the full pulse window in B1913$+$16.
Note the nonzero $L/I$ (solid) and 
the maxium $|V|$ (dotted) 
at the OPM transitions. The result was obtained for 
$\lpk=40^\circ$, $\sigma_{\lag}
= 70^\circ$, and $\sigma_{\psi, in}=30^\circ$.
The arrow in {\bf b} refers to another object
(B1933$+$16) and is discussed in Section \ref{mitraloop}.
}
\label{weismod}
\end{figure}

Therefore,
 the steady motion of $\npsi$ along the horizontal 
axis of the diagram (as caused by the pulsar rotation) translates into the
wavy thick arrow marked in the bottom part of all panels in
Fig.~\ref{snake}. The observed polarisation will be dominated 
by the Stokes parameters recorded along this thick wavy line.
Accordingly, Fig.~\ref{snake} provides $L/I$ (middle panel) and $V/I$ 
(bottom panel) with the same wavy track as in the top panel.

In agreement with the observed properties of B1913$+$16, 
at both OPM transitions ($\psin=\pm45^\circ$) 
the wavy line omits the region of low $L/I$ (bright in the middle panel), 
staying much larger than zero for all the time. 
$|V|/I$ is maximum at the OPM transitions ($\psin=\pm45^\circ$), 
and $V$ changes sign 
in the middle ($\psin=0$), in the region dominated by one polarisation mode. 

A numerical code which convolves Gaussian distributions of $\npsi$ and $\nlag$
gives the result shown in Fig.~\ref{weismod}.\footnote{For each longitude
$\Phi$, i.e.~for a given $\ppk$ and $\lpk$, the code simply runs
over the $\psin$ and $\lag$ loops and calculates the Stokes parameters scaled
by $\npsi\nlag$. A calculation of $\psi$ then makes it possible
 to gather the Stokes 
in separate arrays indexed by $\psi$ and $\Phi$.}
It is obtained for 
 $\lpk=40^\circ$, $\sigma_{\lag}=70^\circ$ and $\sigma_{\psi,in}=30^\circ$ 
(parameters selected to obtain a rough by eye fit).
The value of $\ppk$ changes linearly as defined by the horizontal axis. 
This is likely unrealistic, but appears sufficient to reproduce the
observed polarisation approximately  
(the full length
of the $\ppk$ axis must be cast onto the pulse window of B1913$+$16).
Modelled single pulse PA distributions 
(grey patches in Fig.~\ref{weismod}) 
have the familiar form of OPM bands observed in other pulsars. 
In the intermode region 
(near $\ppk=\pm45^\circ$) 
these bands overlap in $\Phi$, because one wing of $\npsi$ distribution 
extends across the intermode value. 
Similar result is obtained for the $\nlag$ 
distribution shown with the thin line on the left margin 
of Fig.~\ref{snake}. This is because in both cases one  
wing of $\nlag$ reaches the modal points at $\lag=90^\circ$, and $\lpk$ 
is close to zero. 
Even if this last distribution is made symmetric 
around $\lag=0^\circ$, the PA would not change much, but $V$ would vanish.

The direction of the step-wise PA variations 
in B1913$+$16 (Fig.~\ref{weismod}) is determined by the steady motion of 
$\npsi$ towards larger $\psin$ with the increasing $\Phi$. The direction 
of OPM transition is not accidental in the model: within the modal 
transition the derivative $d\psi/d\Phi$ has the same sign as $d\ppk/d\Phi$. 
This tells us that $\ppk$ increases monotonically within the full pulse
window of B1913$+$16, as marked in Fig.~\ref{snake}.

Although no RVM effect is included in the model, the PA curve is fairly 
similar to the observed one. The RVM PA (which may be considered constant  
at this stage of calculation) can be imagined 
as a perfectly horizontal line at $\psi=0$. It can be seen that neither the
average PA nor the grey single-pulse PA track follow the horizontal line. 
This slope is introduced by the steady motion of $\npsi$ across the natural 
mode at $\psi=0$. 
Therefore, in B1913$+$16 the slope of RVM PA is likely biased by the mode 
coherency effects, 
and is unlikely to be determined through the RVM fitting, 
unless a precise model of
the coherent mode combination is included as a part of the RVM fitting
procedure. 
Unlike in Fig.~\ref{weismod}, 
the full span of the observed PA 
($195^\circ$, as noted by Weisberg \& Taylor 2002) exceeds 
$180^\circ$, which is
the maximum possible value for the equatorward viewing geometry. 
This slight vertical extension is likely caused by the RVM, which
only partially contributes to the PA slope observed in the middle of the
profile.

\subsection{Other numerical examples}

\begin{figure}
\includegraphics[width=0.48\textwidth]{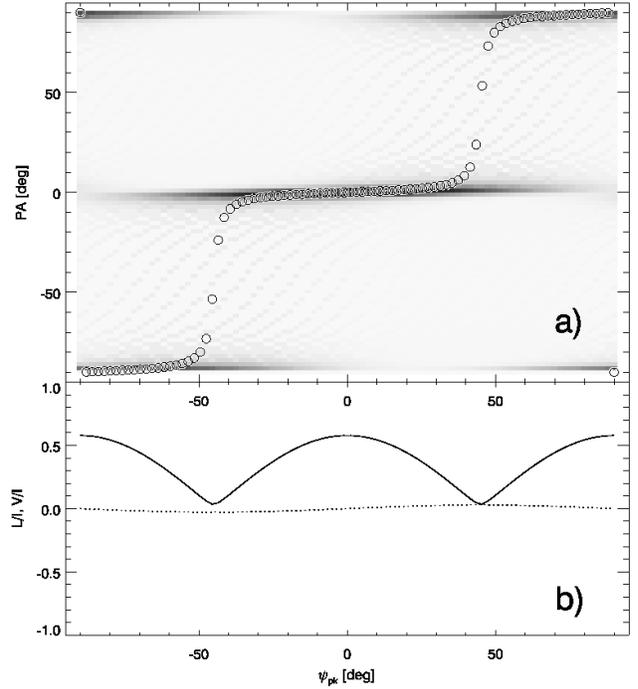}
\caption{Polarisation characteristics for $\lpk=40^\circ$, $\sigma_{\lag}
= 130^\circ$, and $\sigma_{\psi, in}=30^\circ$. The wide $\nlag$ reaches the
modal points at $\lag=90^\circ$ and makes the OPM tracks narrow. The wings of
$\nlag$ extend into both the negative and positive $\lag$, 
therefore, $V$ is suppressed. 
}
\label{try1}
\end{figure}

If the $\nlag$ distribution is very wide ($\wlag\ga\pi$) or centered close to the modal points
($\lag\approx90^\circ$) then thin well-defined bands of both 
OMPs extend for most 
of the profile at a longitude-dependent strength ratio (see Fig.~\ref{try1}). 
This is because the modal points 
on the $P(\psi,\lag)$ plane are always within the reach of $\nlag$ and keep
to be selected by $\nlag$ for any $\ppk$. 
In Fig.~\ref{try1}
$V/I$ is close to zero because 
$\lpk=40^\circ$ is much smaller than $\sigma_{\lag}=130^\circ$ (symmetry of $\nlag$ with
respect to $\lag=0$ suppresses $V$). $L/I$ has 
the profile of $|\cos{2\ppk}|$ with 
deep minima at sharp
OPM jumps.\footnote{$L/I$ has the profile of $|\cos{2\ppk}|$, because 
the input signal of amplitude $E$, linearly polarised at an angle
$\ppk$ with respect to 
one of the natural modes, produces two waves of amplitudes 
$E_x = E\cos\ppk$, and $E_y = E\sin\ppk$.
If $\nlag$ is very wide, i.e.~the waves combine incoherently, then 
$L/I = |E_x^2 -
E_y^2|/(E_x^2+E_y^2)=|\cos(2\ppk)|$, i.e.~$L/I$ vanishes four times 
per a single revolution of $\ppk$ (each time when the projected components 
are equal).}
As soon as $\sigma_{\psi,in}$ is increased above $\sim\negthinspace50^\circ$, 
both sharp modes of similar 
amplitude are present at all $\Phi$ and there is a nearly complete
depolarisation ($L/I=V/I\approx 0$). Overall, very wide distributions tend
to depolarise for obvious reasons, however, the wide 
$\npsi$ suppresses both $L$ and $V$ 
(at whatever width of $\nlag$), whereas the wide $\nlag$ suppresses 
$V$, but $L$ only at the modal transitions.

Moderately wide $\npsi$ can 
produce broad clouds of PA centered at the modal values, 
with the average PA slowly traversing from one mode to another. 
The transition may occur at a small, but non-vanishing $L/I$ and $V/I$. 
Effects of this type are often observed (e.g.~B0823$+$26 in Fig.~12 of MAR15, 
B2110$+$27 in Fig.~19 therein) and occur when $\npsi$ is crossing the
intermode 
and the widths of the distributions are moderate ($\sigma_{\psi,in} 
\sim45^\circ$, $\lpk\sim45^\circ$,
$\sigma_{\lag}\sim45^\circ$). Otherwise, i.e.~for a large $\wlag$,  
$\nlag$ can reach the modal points on the lag-PA plane, and the observed
modal tracks become narrow.

\subsection{Other pulsars}

\begin{figure}
\includegraphics[width=0.48\textwidth]{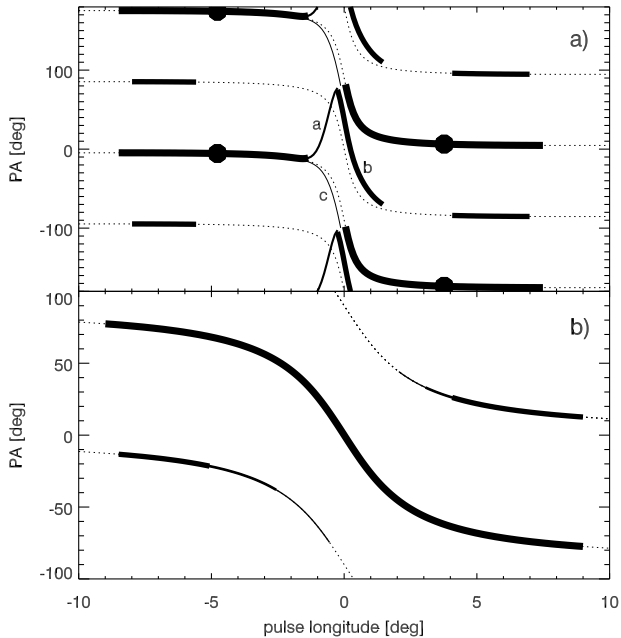}
\caption{Generic examples of PA behaviour in M-type pulsars ({\bf a})
and D pulsars ({\bf b}). In both cases short segments (patches) 
of the secondary mode are present in the profile periphery. 
Panel {\bf a} is based on Fig.~1 in SRM13 and presents the complicated core 
PA distortion in PSR B1237$+$25. The upper branch of it (marked `a' and `b')
is  interpreted in Section \ref{core}. The bullets at 
$\Phi=-5^\circ$ and $4^\circ$ mark the observed 
widening of the primary PA track. Dotted lines present the RVM model assumed
in this paper. 
}
\label{proto}
\end{figure}

Two prototypical single pulse PA scatter plots 
are schematically presented in Fig.~\ref{proto}.
The top one, based on B1237$+$25 (Fig.~1 in SRM13), shows the PA of an 
M-type profile (with multiple components, Rankin 1983; Backer 1976)
\nct{ran83, bac76} 
and with a complicated PA distortion at the profile center. 
Similar distortions,  
 associated with 
the core component, are also observed in B1933$+$16 and B1857$-$26.
Double `conal' profiles (Fig.~\ref{proto}b), on the other hand,  
exhibit the well defined S-shaped PA swing following a single OPM 
across 
the full profile, with roughly symmetric 
patches (short fragments of a PA track) 
of the secondary OPM in the profile peripheries. As shown in
Fig.~\ref{proto}a, similar orthogonal PA patches also appear in the pulsars with the complex 
core properties (e.g.~B1237$+$25). Unfortunately, in these objects
 the complexity of the central PA distortion often makes it difficult  
to consistently identify a single OPM throughout the whole pulse window. 
In B1933$+$16 the loop-like PA distortion converges back at the original
(strong) mode. 
In B1237$+$25, however, the core PA loop spreads from the primary mode,  
but appears to converge at another, secondary 
mode, as identified by the short PA patch marked in Fig.~\ref{proto}a. 
Moreover,  
Mitra \& Rankin (2008) note that flat sections of the PA curve  
observed in B1857$-$26 imply equatorward viewing geometry for which 
the full span of the RVM PA cannot exceed $\pi$, contrary to what seems to
be 
observed, if the RVM is consistently attributed to the primary (strong) mode. 
The authors were then tempted to present a PA fit based on 
`unsavory assumption' that the primary (brighter) mode on the leading side
continues into the `patch mode' on the right.

In principle, however, such change of mode identity (with the RVM 
continuing from the primary into the secondary mode) occurs naturally 
in the CMA model, whenever $\npsi$ moves across $45^\circ$ and starts to
mostly contribute to the other polarisation mode. 
This is because for a phase lag
increasing from zero, the lines of fixed-$\psin$ in the lag-PA diagram (Fig.~\ref{expla}) 
diverge upwards for $\psi_{in}\ga45^\circ$, whereas they diverge
downwards for $\psi_{in}\la45^\circ$. Thus, at an intermode 
the strength of the modes is exchanged and  
the primary mode becomes the patch mode. Such inversion of the 
mode amplitude ratio is also explained in Fig.~\ref{ellipses}a. 
This possibility allows one to construct a naive polarisation model
of pulsars such as B1237$+$25 and B1857$-$26, which resembles the generic
case shown Fig.~\ref{proto}a.

\subsubsection{Tentative discussion of polarisation for pulsars with
 complicated core emission}

\begin{figure}
\includegraphics[width=0.48\textwidth]{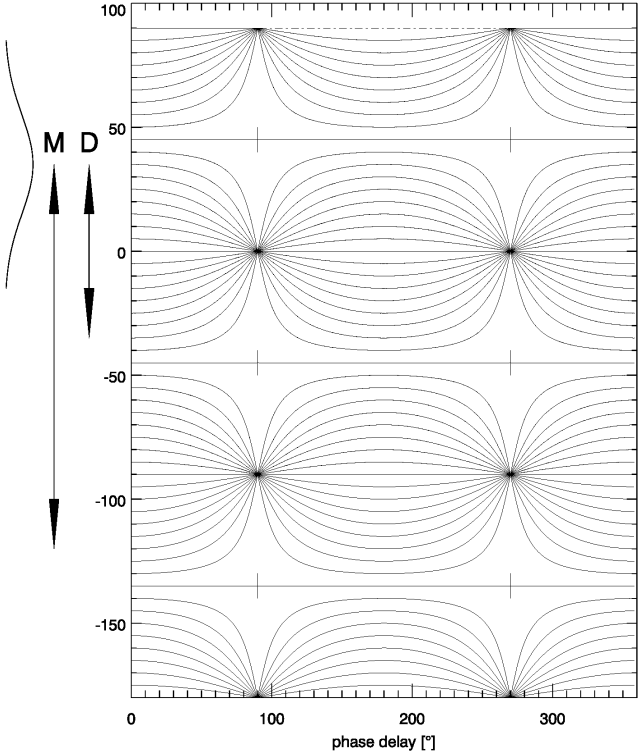}
\caption{Simplest model of M and D pulsar polarisation 
assumes that the $\npsi$ distribution moves within the range indicated by
arrows. In the case of M pulsars this involves a single  passage through 
the $-45^\circ$ intermode, which implies the inversion of modal flux ratio
in the profile center (thus the replacement of the mode identification).
}
\label{expla}
\end{figure}

At the leading edge of the profile of B1237$+$25, 
the $\npsi$ distribution may be thought 
to be located not too far from $45^\circ$ (as marked 
on the left margin of Fig.~\ref{expla}), since both OPMs are observed 
(the primary mode and the secondary or patch mode) and $L/I$ is about 
50\%. This is caused by a 
wing of $\npsi$ reaching across $\psin=45^\circ$. 
The observed zero value of $V$, generally expected when $\npsi$ is
displaced from the modal value ($\ppk=0$ or $\pm90^\circ$), 
might be justified by assuming that 
the 
$\nlag$ distribution is symmetric and centered at $\lag=0$.
With increasing longitude, $\npsi$ moves away from $45^\circ$ since 
the patch mode disappears, whereas the 
primary mode nearly totally dominates the observed flux ($L/I\approx90\%$). 
This corresponds to the peak of $\npsi$  crossing through $\psin=0$
 in Fig.~\ref{expla}.
Before the core is reached from left 
(at $\Phi\approx-2^\circ$ in Fig.~1 of SRM13), 
$L/I$ starts to quickly decrease, because  
one wing of $\npsi$ (the bottom wing in Fig.~\ref{expla}) 
extends across $-45^\circ$. 
This time no patch of the secondary mode appears, but this
might be explained by arguing that 
both modes are simultaneously present in individual samples 
of single pulse emission.
 At the central loop-like distortion 
$\npsi$ would have to cross $-45^\circ$, which produces the OPM transition
and changes the mode illumination, thus replacing the identity of modes
(the bright mode now follows another RVM track offset by $90^\circ$).
Then $\ppk$ could continue decreasing (as marked with the M arrow in 
Fig.~\ref{expla}) and the pattern 
would repeat in a reversed order, 
i.e.~$L/I$ would increase, and $\npsi$ would approach
$\psi_{in}=-90^\circ$, 
since again one mode
dominates in the data on the right hand side of the core, 
where $L/I$ is very high. Finally $\npsi$ approaches the intermodal 
$\psin=-135^\circ$, and the patch of the now-secondary mode lights up.
In this scenario, within the profile window of `complex core pulsars', 
the peak of $\npsi$ moves by nearly $180^\circ$, a value 
that may be associated with a sightline passing very close 
to the magnetic pole.

This interpretation may seem to be consistent with  the upper branch of the PA
loop in B1237$+$25, which separates from the primary mode at $\Phi=-0.8^\circ$,  
but converges at the patch mode PA at longitude $\Phi=1.5^\circ$ 
(see Fig.~1 in SRM13). Had it been correct, such model would imply
that the PA curve of B1237$+$25
should be fitted with 
the RVM which follows the primary mode on the left hand side, but   
continues into 
the patch mode on the right hand side of the profile.
The `unsavory assumption' of Mitra \&  Rankin (2008), made for
B1857$-$26 (Fig.~2 therein) then may be considered a realistic possibility
for pulsars such as B1237$+$25 and B1857$-$26.
With the odd number of intermode crossings within the core, 
the distinction of the primary mode and the secondary mode (patch mode) 
is meaningless 
if it is supposed to reflect a single RVM track through the entire pulse
window. However, this picture ignores some details of the observed core PA
distortion, and the geometric origin of the $\npsi$ motion, so it  will be 
rectified further below.


\subsubsection{Tentative polarisation model for classical D-type pulsars}

Unlike the `complex core' pulsars, pulsars such as B0301$+$19 and B0525$+$21 
exhibit textbook PA variations that seem to stay in a single RVM track 
through the whole pulse window. Interpretation 
based on the motion of $\npsi$ (Fig.~\ref{expla})
 implies that $\npsi$ moves within much smaller interval 
in these objects. 
On the leading edge of their profiles,  
one wing of $\npsi$ must extend across $\psi_{in}=45^\circ$, since 
the `patch mode" and low $L/I$ are observed there. In the profile center 
 $\npsi$ 
crosses zero, which would explain why $L/I$ is larger in the central region
 (which is counterintuitive, since two profile components that
overlap in pulse longitude, may be expected to depolarise each other 
at the center).
At the trailing side $\npsi$ would approach $-45^\circ$, again feeding 
the existence of unequal amounts of two OPMs. 
The range of traversed $\psi_{in}$ is smaller than $90^\circ$, which
may seem consistent with  a 
passage of sightline at a larger distance from the magnetic pole. 
The primary and secondary modes do not have their identity replaced  
in D-type pulsars, 
so the traditional RVM fitting may be applied. 

This said, it must be noted that alternative interpretation of D pulsars 
is possible at least for $L/I$,  because $L/I$ is symmetric with respect to $\psin=0$ 
on the lag-PA diagram. Accordingly,  
$\npsi$ may move from the vicinity of $45^\circ$ towards some value
close to zero, and then retreat towards the original position. 
In this case $\psin$ changes nonmonotonically and 
has a minimum in the middle of the profile. The resulting linear 
polarisation is 
nearly identical to the case with the steadily increasing $\psin$. 
Section \ref{picture}, which derives the motion of $\npsi$
from the magnetospheric geometry, implies yet a different variations of 
$\ppk$.

\subsection{Core PA distortion in B1933$+$16}
\label{mitraloop}

The core PA  distortions look like fast modal transitions which 
have got reversed, or failed to be completed.
As can be seen in Fig.~\ref{modes}, any manipulations with $\nlag$ 
(displacements of $\lpk$ along the $\lag$ axis) do not allow us 
to leave the single mode space between the consecutive intermodes.
On the other hand, 
a single passage 
of $\ppk$ through one of 
the intermodes (at $\pm45^\circ$ or $\pm135^\circ$) 
  produces both the orthogonal transition of PA 
and a change of the modal amplitude ratio, 
thus redefining which mode is primary
 and which is secondary.

Still, such replacement of the mode illumination 
requires an odd number of intermode crossings, whereas 
the core PA loops have
the `up-and-down' form, which suggests an even number 
(on the way to another mode and back).
Even in B1237$+$25 the odd-numbered 
transition of the core PA (from the primary to the secondary mode)
 is followed
by one more PA transition back to the primary mode, at a trailing-side 
longitude where the
core emission seems to cease and subdue to the peripheric emission.
 
In B1933$+$16 the core loop opens and closes at the same (say
primary) mode, so it will now be interpreted in terms of the 
reversed mode transition with no mode identity replacement, as 
caused by nonmonotonic variations  of $\ppk$. 
On the left side of the PA loop in B1933$+$16 
(see Fig.~1 in MRA16) 
one polarisation mode
dominates which means that $\ppk\sim0$ and 
the observed $\psi$, $V/I$, and $L/I$ can be represented 
by the values at the median line of Fig.~\ref{weismod} (assuming similar 
$\nlag$ distribution for B1933$+$16 and B1913$+$16). 
Let us assume that within the loop of B1933$+$16 
$\ppk$ changes with $\Phi$ nonmonotonically, 
in the way which is presented by the backward-bent arrow
in Fig.~\ref{weismod}b. This qualitatively reproduces the major 
observed properties of the loop: 
while moving leftward $L/I$ decreases, $|V|/I$ increases, and the average PA 
almost makes the full OPM transition.
Shortly after passing through the minimum in $L/I$ (at $\ppk=-45^\circ$), 
$\ppk$ starts to
increase, immediately passing again through the $L/I$ minimum, and 
$|V|$ starts to drop. 

\begin{figure}
\includegraphics[width=0.48\textwidth]{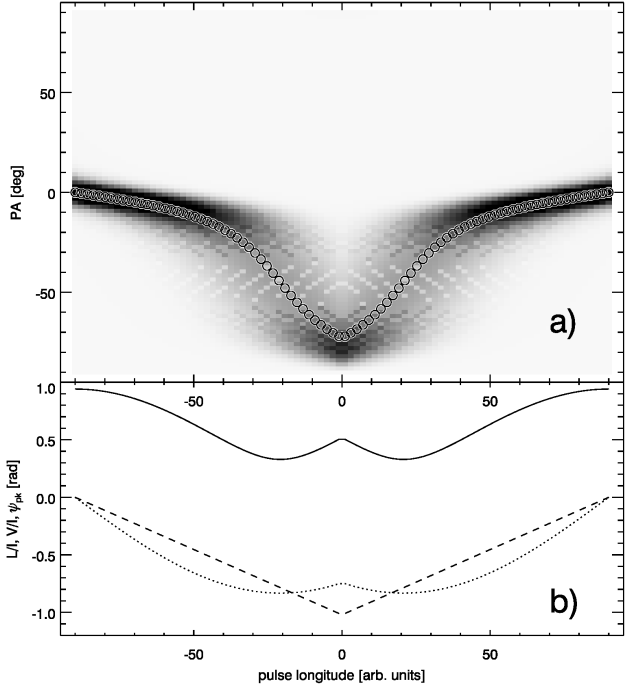}
\caption{A model for the V-shaped PA distortion observed at the core of
B1933$+$16 (cf.~Fig.~1 in MRA16). 
Despite $\ppk$ following the primitive linear model marked with the dashed
line in {\bf b}, the twin minima in $L/I$ (solid) and the large negative $V$ 
(dotted) are similar to those observed.
}
\label{loop}
\end{figure}

Numerical simulation of such case is shown in Fig.~\ref{loop}, 
where simple linear 
changes of $\ppk$ were asumed ($\ppk$ in radians is shown in the bottom panel 
 with a dashed line). 
Despite such linear 
variations of
$\ppk$ 
are far from those expected at an even-numbered intermode crossing in pulsar
magnetosphere 
(see Section \ref{picture}) they produce polarisation similar to that
observed. In particular, 
the modelled minima in $L/I$ have the characteristic  
twin-like look as in the data: they are close to each other, 
and connected with only a slightly increased $L/I$ in between them.
The sharp tip at the center of $L/I$ and the slanted PA on both sides of the 
 V-shaped PA feature 
(Fig.~\ref{loop}) are artifacts of the linear 
$\ppk$. 

The V-shaped distortion of Fig.~\ref{loop} does not extend beyond
$\Delta\psi = 90^\circ$ hence it does not have the loop-shaped form. 
The observed shape may be caused by a more complicated structure 
of $\npsi$ than the
Gaussian form assumed above. 
However, the observed loop can also be interpreted  
purely in terms of $\nlag$ 
displacement, with a fixed 
$\ppk\sim45^\circ$ within the entire feature. The PA curve
distortion is then bidirectional, i.e.~the primary PA track bifurcates 
into a loop. 

\subsubsection{Longitude-dependent $\nlag$ and the loop of B1933$+$16}

\begin{figure}
\includegraphics[width=0.48\textwidth]{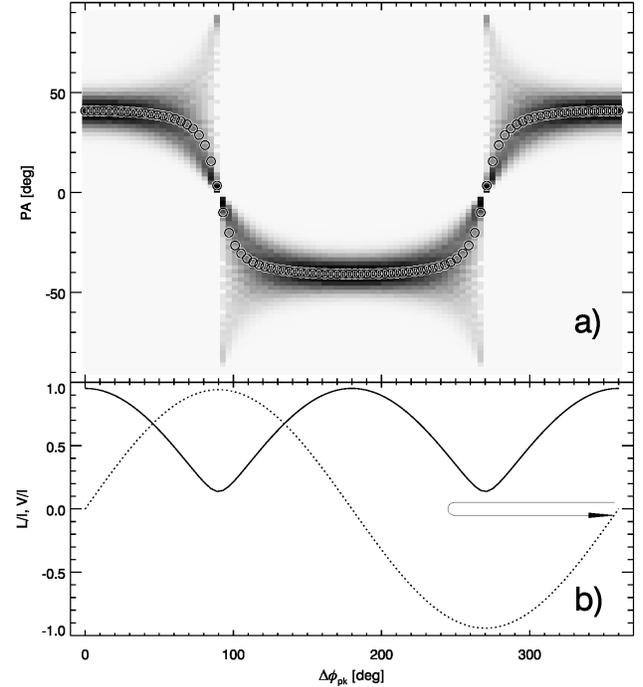}
\caption{Polarisation characteristics calculated for a fixed $\npsi$ 
($\ppk=41^\circ$, $\sigma_{\psi,in}=5^\circ$) with a narrow 
$\nlag$ ($\sigma_{\lag}=15^\circ$) centred 
at the values marked on the horizontal axis. Nonmonotonic 
changes of $\lpk$, marked with the retreating arrow in {\bf b}, 
produce the result shown in Fig.~\ref{petla2}.
}
\label{petla}
\end{figure}

If $\npsi$ is fixed at a value close to $45^\circ$ ($\ppk=41^\circ$ in 
Fig.~\ref{petla}) then the loop-like bifurcation of the PA track 
can be interpreted through the motion of $\nlag$ along the $\lag$ axis. 
When the full $\lpk$ interval ($360^\circ$) 
is cast onto some $\Phi$ interval, 
then the loop of Fig.~\ref{petla} is formed. 
The loop appears because the PA track nearly follows the disconnected
sections of the broken intermodal PA (thick in the bottom panel of
Fig.~\ref{modes}). 
Since $\ppk < 45^\circ$, with the increasing $\lag$ 
most of input power (in $\npsi$) is displaced down, 
towards the lower intermode, whereas a part of the $\npsi$ wing (where 
$\psin > 45^\circ$) follows the upper branch of the loop. 
At the proper modes ($\psi=0$ or $90^\circ$, $\lag=90^\circ$ or
$270^\circ$) the PA track is strongly enhanced\footnote{The blank 
breaks at the proper modes of Fig.~\ref{petla} 
result from a limited resolution of the calculation.}
and narrow. This results from the fixed orientation of the assumed 
intervening polarisation basis, which in reality can fluctuate. This 
would reduce the modal enhancement and make the PA track at the modal points 
wider.

\begin{figure}
\includegraphics[width=0.48\textwidth]{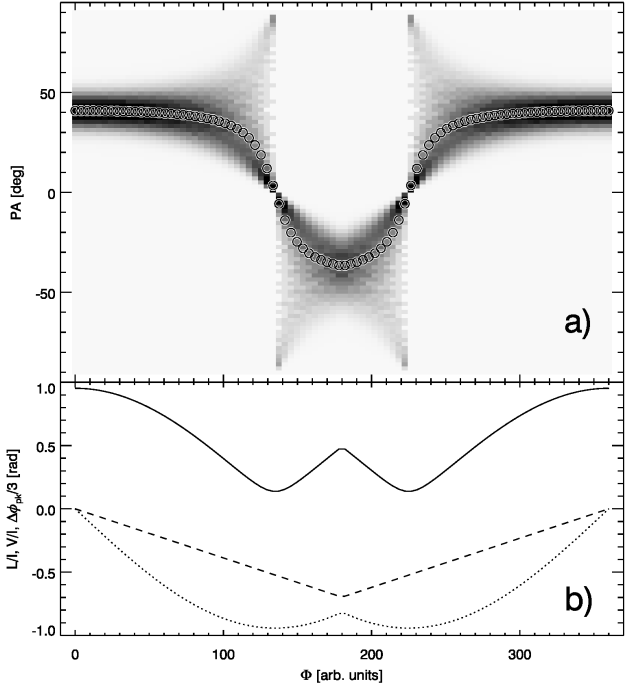}
\caption{Polarisation model for the PA loop of B1933$+$16. $\lpk$ decreases
linearly to $-120^\circ$ (at $\Phi=180$), then increases to $0$ 
(to maintain consistent layout, the dashed line in {\bf b} presents  
$\lpk/3$ in radians). The result roughly reproduces 
the loop-shaped PA track, the  
double minima in $L/I$ and the negative $V$ observed 
at $1.5$ GHz. The same parameters
as in Fig.~\ref{petla} are assumed.
}
\label{petla2}
\end{figure}

Fig.~\ref{petla} does not reproduce the $1.5$ GHz PA loop observed 
in B1933$+$16, 
because $V>0$ on the leading side (see.~Fig.~1 in MRA16). 
However, a nonmonotonic change of 
$\lpk$ within the loop such as marked with the backward-bent arrow 
in Fig.~\ref{petla}b, produces the result of Fig.~\ref{petla2},
which is qualitatively consistent with data. In particular, the twin
minima in $L/I$ appear in the middle of the intermodal transitions 
(i.e.~coincident with the proper modes) and $V/I$ stays negative throughout
the loop. The maxima of $|V|$ coincide with the minima in $L/I$, as observed
on the leading side of the loop in B1933$+$16.

\begin{figure}
\includegraphics[width=0.48\textwidth]{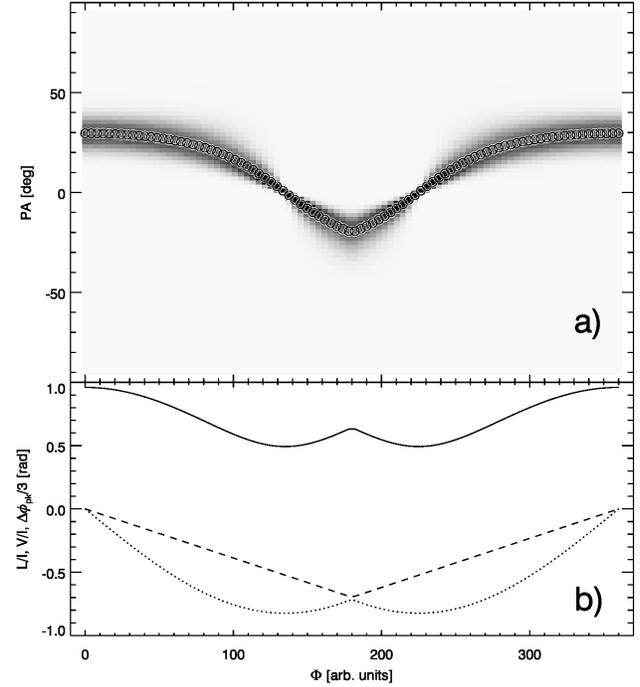}
\caption{Polarisation model for the PA distortion of B1933$+$16, as observed 
at $4.5$ GHz. The value of $\ppk=31^\circ$ is further from the intermode
than at $1.5$ GHz, 
hence the top part of the loop disappears and 
the PA curve amplitude decreases. $L/I$ is larger, whereas $V/I$ -- smaller
in comparison to the case of $1.5$ GHz.
}
\label{petla3}
\end{figure}

At a higher frequency $\nu=4.5$ GHz the upper branch of the loop disappears 
(Fig.~1 in MRA16), the amplitude of the V-shaped average-PA distortion 
decreases, $L/I$ increases, whereas $V/I$ decreases.
All these properties are qualitatively reproduced 
when the misalignment of $\npsi$ from
the intermode is increased at the larger $\nu$. This can be seen in 
Fig.~\ref{petla3} which presents the result for $\ppk=31^\circ$.

The lag-based intermodal split of Fig.~\ref{petla2} 
is then a successful model, capable of reproducing several observed 
properties at both frequencies. The nonmonotonic increase and drop of
$|\lpk|$ suggests a passage of a signal through a stream or stripe of
intervening matter, whereas the change of $\ppk$ with $\nu$ 
could be associated with the $\nu$-dependent emission altitude 
or PLR altitude.
However, the model is more complex than 
the $\ppk$-based model of Fig.~\ref{loop}. Namely, if the primary
mode observed just left of the loop in B1933$+$16
 results from the intervention of some PLR polarisation basis, then 
another intervening basis, offset by about $45^\circ$, 
is required within the loop.
Therefore, in the following material I rather focus on the motion 
of $\npsi$ and on the subject of intermode 
crossing in the central part of the profile.

\subsection{Magnetospheric picture}
\label{picture}

\begin{figure*}
\includegraphics[width=0.78\textwidth]{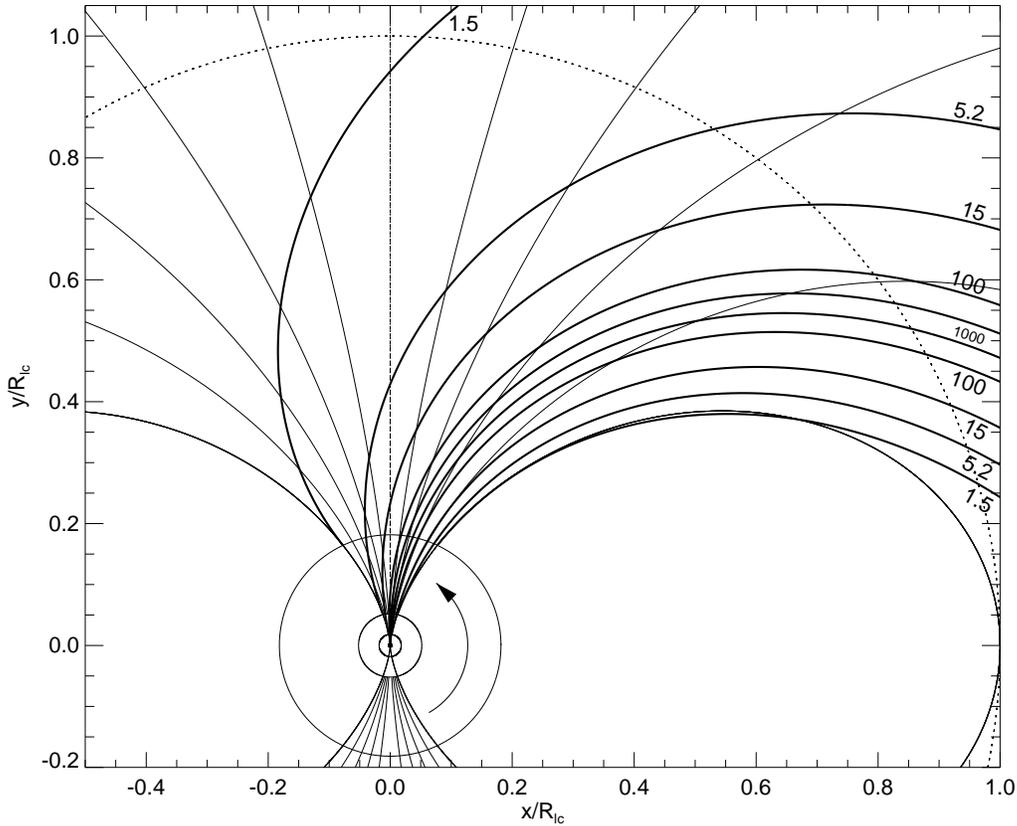}
\caption{Calculated trajectories of photons emitted 
from the polar cap rim and shown in the 
reference frame that corotates with a pulsar 
in the direction indicated by the
arrow. The photons move in the plane of rotational 
equator and the magnetic dipole is tilted orthogonally with respect
to the rotation axis.
Pairs of trajectories (one leaving the leading side, another -- the trailing
side of the polar cap)
 correspond to different rotation periods as labelled in ms on the right. 
The size of neutron stars,  
centered at $(0,0)$ and spinning at these periods, is 
scaled as appropriate fraction of the light cylinder radius $\rlc$. 
Note the backward bending of the emitted beams, which effectively 
dislocates the
high-altitude dipole axis towards the leading side and out of the profile. 
}
\label{farah}
\end{figure*}

Previous sections suppose and imply that the input PA distribution 
rotates with respect
to some intervening polarisation basis.
A possible interpretation of this is to place the intervening basis at the 
 PLR and  orient it along the sky-projected
 local magnetic field.
The maximum of $\npsi$, on the other hand, is attributed 
to the projected direction 
of magnetic field line planes in a low-altitude emission region.
When the emitted beam is propagating 
 outwards through the rotating magnetosphere, 
it bends backwards in the reference frame which 
corotates with the star (Fig.~\ref{farah}). 
With the increasing radius $r$, dipole axis  moves out of the beam in the
forward direction.

\begin{figure}
\includegraphics[width=0.48\textwidth]{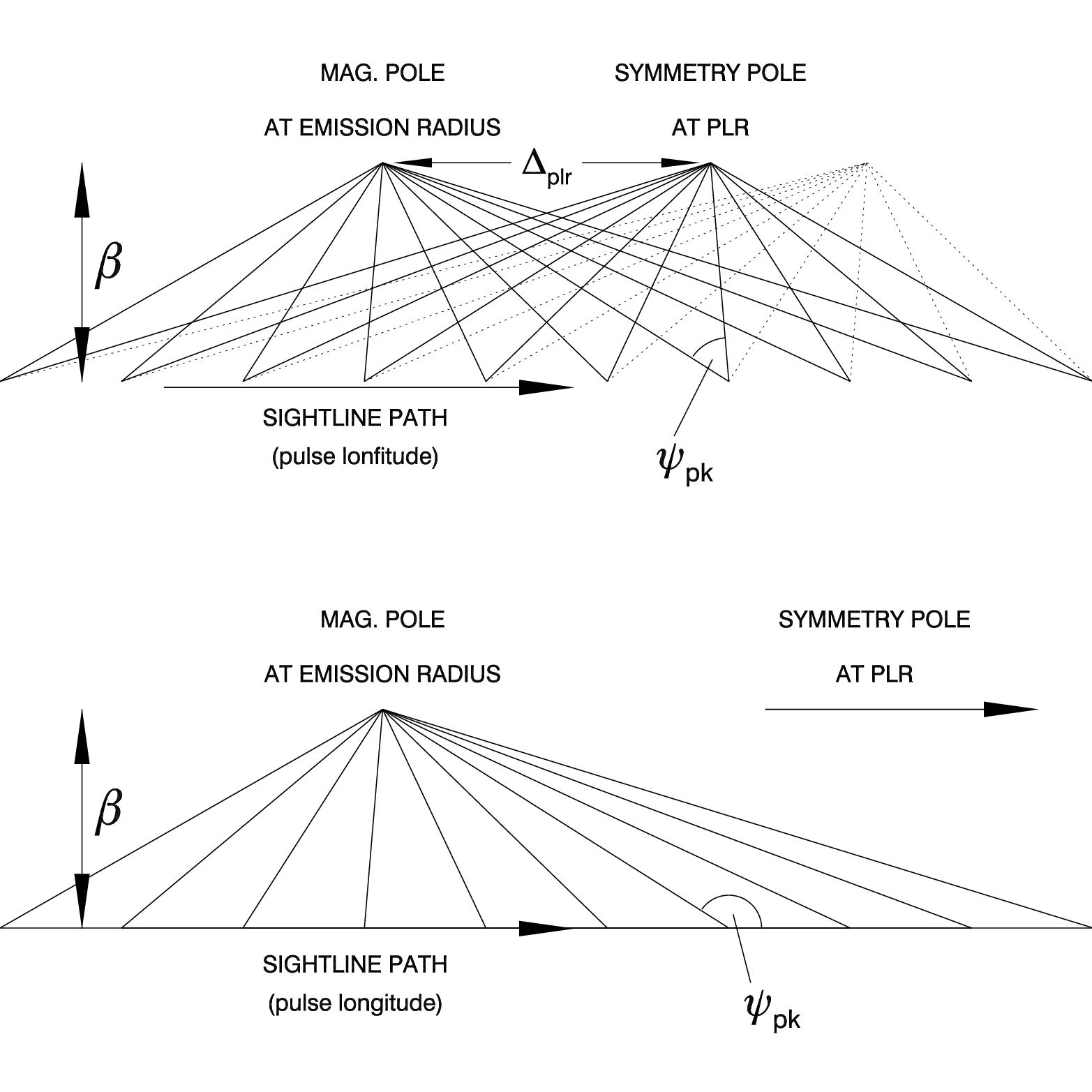}
\caption{Top: orientations of polarisation directions recorded by the line of
sight while moving across the pulse window. 
Solid lines emerging from the left pole mark the  
sky-projected directions of dipolar 
$B$-field at the radial distance of emission $r_{em}$. 
Solid lines emerging 
from the right pole mark the polarisation direction 
of the intervening polarisation basis and supposedly reflect the 
projected structure of high-altitude electron trajectories at the PLR. 
The incident
polarisation angle $\ppk$ is measured between the lines of both structures, 
at the location selected by the line of sight. Dotted line structure 
shows the effect of increased altitude difference between $r_{em}$ and PLR.
A low-$r_{em}$ circular emission beam 
should be centered at the left pole. 
Bottom: same as above but for extreme separation of both line structures.
}
\label{druty}
\end{figure}

This suggests the structure of two displaced $B$-field line patterns
(Fig.~\ref{druty}), one of
which represents the symmetry at the low altitude of emission (radial
distance $r_{em}$), whereas the other 
 -- at the PLR. Lines in both these patterns rotate at different rate
while they are probed by the horizontally 
passing line of sight. 
As marked in the figure, the input polarisation angle $\psi_{in}$ 
is determined by the angle at which 
 the lines in these patterns cut each other at a point selected by the line of
sight.

The material of previous sections implies that the average PA observed at a
given $\Phi$  
usually coincides with the OPMs of the intervening basis, hence the observed
RVM curves must be attributed to the PLR. The low-$r$ emission, 
on the other hand, is only backliting the PLR polarisation basis. The
polarisation direction of the low-$r$ emission determines 
the flux ratio of observed modes, 
e.g.~it can produce 
 the orthogonal mode transitions, but otherwise 
does not affect the observed PA value.

Fig.~\ref{farah} suggests that the PLR 
line structure should be positioned on the left hand-side (the leading side)  
of the emitted beam. However, this would move the RVM PA curves 
towards the leading side of profiles, in conflict with most observations
(Blaskiewicz et al.~1991; 
Krzeszowski et al.~2009).
\nct{bcw91, kmg09}
Therefore, below I risk the assumption that dynamics of plasma at the PLR  
 is influenced by noninertial effects of corotation, 
in such a way that 
the effective `$B$-field line' structure\footnote{Actually the structure 
of electron trajectories in the observer reference frame, 
see Fig.~2 in Dyks et al.~(2010), cf.~Blaskiewicz et al.~(1991), Dyks
(2008), and Kumar \& Gangadhara (2012).
\nct{dwd10, d08, bcw91, kg12}}
 is displaced rightwards with
respect to the sky-projected structure of the low-$r$ $B$-field. 

\begin{figure}
\includegraphics[width=0.48\textwidth]{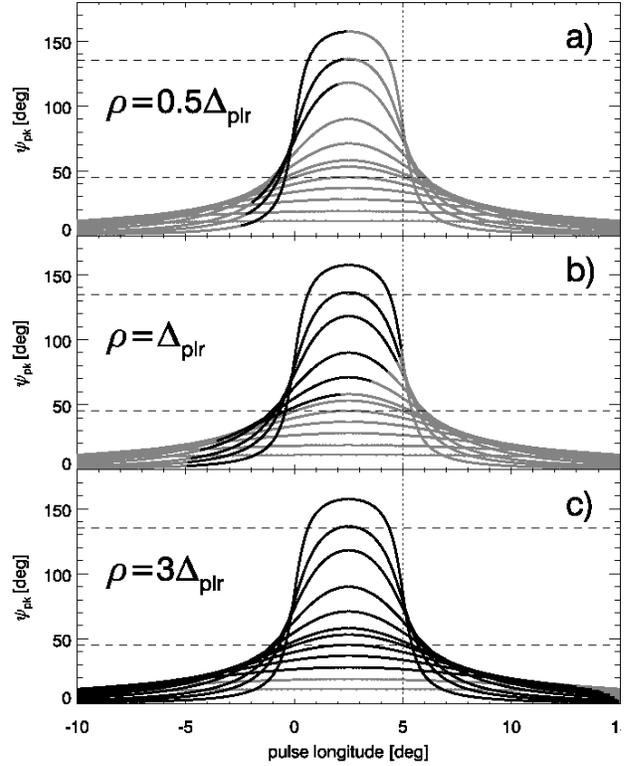}
\caption{Variations of the incident PA $\ppk$ that result from 
the sightline traverse near the magnetic dipole axis, calculated for the
flat geometry of Fig.~\ref{druty}. The pole at $r_{em}$ is located at
$\Phi=0$, whereas the PLR pole is located at $\Phi=\Delta_{plr}=5^\circ$ 
(dotted vertical line).
Different lines, the same in all panels, 
correspond 
to $\beta/\Delta_{plr}=0.1$, $0.2$, $0.3$, $0.5$, $0.7$, $0.9$, $1.0$, $1.2$, 
$1.5$, $2.0$, $3.0$, and $5.0$ (top to bottom). The black portions 
of the lines are detectable if the circular radio beam, 
centered at $\Phi=0$, has the half opening angle $\rho=0.5$, $1$, and
$3\Delta_{plr}$. The different beam extent allows for both monotonic and
nonmonotonic changes of $\ppk$ in the profile, as well as odd or even number 
of intermode crossings at the dashed horizontal lines.
}
\label{szprychy}
\end{figure}

To gain a quick insight into the phenomenon, a flat geometry of
Fig.~\ref{druty} is assumed (instead of the spherical). If the radio 
beam center 
(and the center of an observed pulse profile)
is placed at $\Phi=0$, the 
beam and PLR polarisation angles are:
\begin{equation}
\psi_{bm}=\arctan(\beta/\Phi)
\end{equation}
\begin{equation}
\psi_{plr}=\arctan(\beta/(\Phi+\Delta_{plr})),
\end{equation}
where $\beta=\zeta-\alpha$ is the impact angle i.e.~the angle of closest
approach of sightline to the magnetic pole, and $\Delta_{plr}$ is the angular displacement  
between the poles at $r_{em}$ and PLR. The input PA is equal to
\begin{equation}
    \ppk=\psi_{plr}-\psi_{bm}
\end{equation}
and is shown in Fig.~\ref{szprychy} for a set of $\beta$ values ranging between 
$0.1\Delta_{plr}$ and $5\Delta_{plr}$.

As might be expected, the curves keep close to $\ppk=0$ 
at large $|\Phi|$ and 
are symmetric with respect to the midpoint between 
the beam center at $\Phi=0$ and the PLR symmetry point 
at $\Phi=\Delta_{plr}=5^\circ$. 
For a nearby pole passage, $\ppk$ follows the upper curves (hereafter I
refer to the curves' position at the
midpoint $\Phi$), which extend vertically for a large fraction of $\pi$.
Some of them cut the intermodes (at $45^\circ$ or $135^\circ$) 
twice or four times. For $\beta = 0.2\Delta_{plr}$ (second line from top) 
the top intermode (at $\ppk=135^\circ$) 
 is approached from below with a subsequent retreat. For a more distant polar 
passage, $\psin$ can approach the $45^\circ$ intermode from below, 
or stay close to zero throughout the profile. 
In the extreme case of  $\beta \ll
\Delta_{plr}$, the geometry shown in the bottom of Fig.~\ref{druty} holds, with 
$\ppk\approx\pi-\psi_{bm}$, i.e.~$\ppk$ increases monotonically. 
In the last case the intermodes (at $\ppk=45^\circ$ and $135^\circ$)
 are cut twice
at two longitudes located symmetrically on both sides of $\Phi=0$. This
resembles the case of B1913$+$16. If the location of $r_{em}$ (or PLR) 
changes with 
the observation frequency $\nu$, then $\Delta_{plr}$ can change 
which leads to an increase (or a decrease) of $\ppk$ consistently at all 
longitudes (dotted lines in Fig.~\ref{druty}).

The black portions of lines in Fig.~\ref{szprychy} mark which 
parts of the $\ppk$ curves are detectable, if the radio emission is limited
to a dipole-axis-centred cone with a half-opening angle $\rho=0.5$, $1$, and
$3\Delta_{plr}$ (panels a, b, and c, correspondingly). 
As can be seen in Fig.~\ref{szprychy}, the detectable variations of $\ppk$ 
can either be  monotonic or have a maximum located asymmetrically in
the profle. When the PLR displacement is small ($\Delta_{plr} \ll
\rho$, Fig.~\ref{szprychy}c), fast symmetric changes of $\ppk$
 are located in the center of the profile.

\begin{figure}
\includegraphics[width=0.48\textwidth]{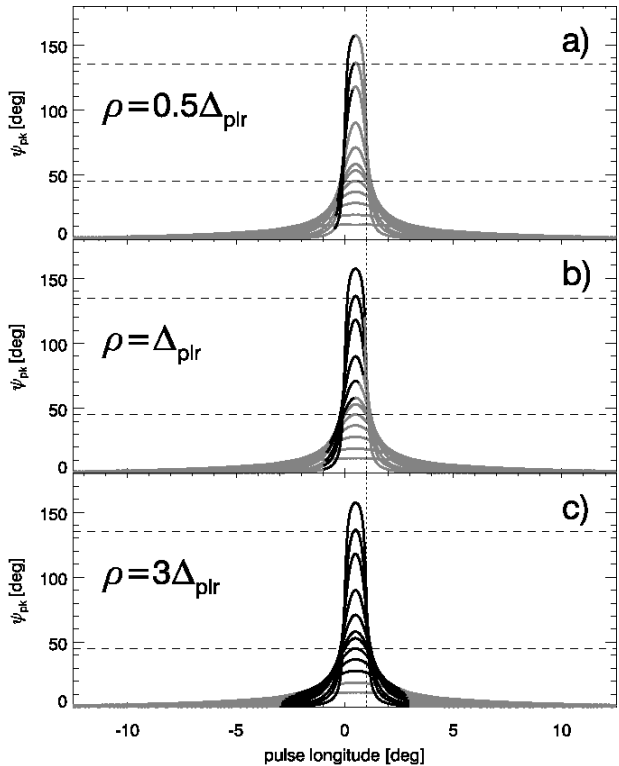}
\caption{Same $\ppk$ profiles as in the previous figure, but calculated 
for a smaller PLR misalignment $\Delta_{plr}=1^\circ$.
Hence this case corresponds to a five times smaller 
beam radius $\rho$ and detectability of intermode crossings 
 refers to the narrow central core component.}
\label{szbis}
\end{figure}

Even for the large displacement of the PLR line structure
($\Delta_{plr}=5^\circ$ in Fig.~\ref{szprychy}), 
the tendency to cross intermodes near the center of the profile is clearly
visible. If $\Delta_{plr}$ is five times smaller (and equal to $1^\circ$) 
the curves 
are horizontally compressed as shown in Fig.~\ref{szbis}, 
which puts all the intermode 
crossings really near the profile center (in the `core' region).
The passage of sightline which is sufficiently close to the 
pole to cross at least the lower intermode is then less likely. 
A simple calculation
shows that $\ppk(\Phi)$ crosses the lower intermode (at $\psin=45^\circ$)
two times for $\beta<0.5\tan(3\pi/8)\Delta_{plr}=1.21\Delta_{plr}$. 
Both intermodes (at $45^\circ$ and $135^\circ$) are crossed 
when $\beta<0.5\tan(\pi/8)\Delta_{plr}=0.21\Delta_{plr}$ (in which case 
there may be up to four detectable crossings). Thus, to produce a complicated, 
loop-like PA
feature in the profile, the line of sight must pass at about the same
distance from the pole  as $\Delta_{plr}$ (or closer).

When $\beta\ga1.2\Delta_{plr}$, the increasing   
$\ppk$ approaches the lower intermode and drops to the
original low value without the crossing (5th line from bottom in 
Figs.~\ref{szprychy} and \ref{szbis}). 
Moreover, if the angular radius $\rho$ of the emission cone 
is much larger than $\Delta_{plr}$ 
(Figs.~\ref{szprychy}c and \ref{szbis}c) then the number of intermode 
crossings is even, which implies no replacement 
of the mode illumination or identification in the profile peripheries.
Therefore,
the presence of the PA loop under the core component does 
not necessarily mean that the primary mode on one side of the profile 
should continue into the secondary (patch) mode on the other side.
However, for the bottom black cases 
in Fig.~\ref{szprychy}a and b, 
such replacement of the mode identification does occur 
at the lower intermode crossing near the center of the pulse window. 

\subsection{Distribution displacement vs widening}

If the $\ppk$ profiles of Fig.~\ref{szprychy}
are considered as a valid model for the motion of $\npsi$ 
across the lag-PA diagram of Fig.~\ref{modes}, then 
a problem appears 
with the above described interpretation of D-type profiles. Namely, the
symmetric PA patches in the outer profile region (see
Fig.~\ref{proto}b) have been attributed to the proximity of $\npsi$ to 
the intermodes, whereas for the distant polar passage 
$\ppk$ approaches $\sim 45^\circ$ in the middle of the profile. A double 
traverse through $45^\circ$ matches the behaviour of B1913$+$16
(with its `another' mode at the center), but not B0301$+$19, or B1133$+$16.  

The magnetospheric interpretation (Figs.~\ref{druty} and \ref{szprychy})
suggests that $\ppk$ should be close to zero in the profile periphery, 
and closer to $45^\circ$ in the center.
A possible way out 
is to assume that $\beta \gg \Delta_{plr}$ in D type pulsars,
which is consistent with the peripheric viewing geometry. 
Then $\ppk$ follows the lowest curves shown in Fig.~\ref{szprychy},  
staying close to zero throughout the entire profile. 
The $\npsi$ distribution is then pinned at $\ppk\approx0$ and a method 
to produce the peripheric secondary mode is needed. A natural solution is 
to assume that $\npsi$ becomes much wider in the profile periphery.
Both wings of $\npsi$ extend across both intermodes (at $\pm45^\circ$) 
thus producing the peripheric patches of the secondary mode.

This interpretation 
is supported by the single pulse observations of D pulsars,
which demonstrate intense clouds of PA samples filling in the space between 
the peripheric OPM tracks (HR10; Young \& Rankin 2012;
 MAR15). \nct{hr10, yr12, mar2015}
Most importantly, 
in the PA distributions observed at a fixed $\Phi$, 
the modal peaks are bridged with wings that seem to extend at equal strength 
towards both the larger and smaller PA. 
This is a signature of a broadenning 
of the incident $\psin$ distribution, not a displacement.

Similar effects are involved in the peripheric double mode PA tracks
 observed in the M type pulsars. 
The widening of the incident $\npsi$ is revealed by thick PA blobs centered 
at the PA track of the primary mode in B1237$+$25 
(at $\Phi=-4.7^\circ$ and $3.7^\circ$
in Fig.~1 of SRM13, also shown with bullets in Fig.~\ref{proto}a). 
The incident $\npsi$ distribution 
must become wider right at these longitudes. Closer to the profile edge 
the primary mode track 
becomes \emph{apparently} thin again, most likely because $\npsi$ has 
become even wider and the power of the primary mode 
is leaking into the secondary mode.\footnote{Hence only the thin topmost 
part of the primary track leaves a visible trace in the figure.}
It should be noted that diversity of $\lag$ values 
at a given $\Phi$ can spread even 
a single input
$\psin$ value into a wide PA distribution (see Fig.~\ref{modes}), however, 
the phase lags alone usually  
cannot make the secondary mode to appear.\footnote{Unless the intermodes
are misidentified as the natural modes, which might happen for the 
case shown in Fig.~\ref{petla}. 
A widening of
$\wlag$ can also switch on the secondary mode track 
if $\ppk\sim0$ and $\wpsi$ reaches across
$\pm45^\circ$, cf.~Fig.~\ref{45}.}
Therefore, the increase of the observed PA
distribution, noted already by SRM13, likely needs that the intrinsic 
$\npsi$ becomes wider.

The core PA distortions may then be also associated 
with a fast increase of $\wpsi$, as evidenced by the
low $L/I$.  
Fig.~\ref{szbis} shows that $\ppk$ is really likely to shoot up 
in the center of the profile, which directly contributes to the increase 
of $\wpsi$. The fast motion of $\npsi$ and the large $\wpsi$ 
are then positively corellated.
It is then found that the width $\wpsi$ of the incident $\npsi$ distribution 
is an important parameter which can vary strongly across the pulse window 
and influence the PA observed in single pulse data.


\subsection{Polarisation of the core emission}
\label{core}

The tools of previous sections may now be used to interpret 
the polarisation of the core component in PSR B1237$+$25.
 
At $\Phi=-3^\circ$ in the profile of B1237 (see Fig.~1 in SRM13) 
$L/I$ is almost 
equal to $1$ and $V/I$ is negligible which means that 
the observed
radiation is totally confined in a single polarisation mode, with
$\ppk\sim 0$ and $\wpsi$ not extending beyond $\pm45^\circ$. 
This corresponds to the locus $(\psin,\lag)=(0^\circ,45^\circ)$
on the lag-PA diagram of Fig.~\ref{snake}.
While moving towards the core, i.e.~rightwards in Fig.~\ref{snake}, 
at $\Phi\approx -2^\circ$ (in Fig.~1 of SRM13), the observed  
$L/I$ decreases and $V/I$ increases in roughly reciprocal relation, because
$\npsi$ starts to deviate from $\ppk=0$, as implied by Fig.~\ref{szbis} 
(the $\ppk(\Phi)$ curves are more clear in Fig.~\ref{szprychy},
 so below I refer to the third from top case in Fig.~\ref{szprychy}b).
The anticorrelation of $L/I$ and $|V|/I$
is characteristic of the coherently combined modes, 
since the minima of $L/I$ clearly overlap with the maxima of $|V|/I$
in the whole parameter space (compare the locations of the ``oval" contours 
in the two bottom panels of Fig.~\ref{snake}).
At $\Phi\approx -0.9^\circ$, $\npsi$ approaches the intermode crossing 
($\ppk\sim45^\circ$) and 
becomes wider, 
so $L/I$ nearly vanishes,\footnote{In the context of the CMA model,
 the following statement from 
SRM13: `the deep
linear minimum just prior to the core shows that the core and the [adjacent] 
emission represent different OPMs' is understood as a 
statement about components of a single incident polarisation vector, 
i.e.~only involves a single input polarisation mode.}  
 and the top branch of the PA loop detaches from the primary PA track
(or the primary PA track bifurcates, see Sect.~\ref{abnormal}).
 The large width of the $\npsi$
distribution keeps the $L/I$ at a low level throughout the entire core. 
At $\Phi=0$, i.e.~slightly on the left side of the core maximum, 
the peak of $\npsi$ passes through 
the natural mode at
$\psin=90^\circ$ which produces the sign change of $V$, 
and the slight increase of $L/I$ in between the twin minima. 
In Fig.~\ref{proto}a this corresponds to the passage of the `a' branch 
of the core loop through the dotted RVM curve of the secondary polarisation
mode.
Then, while following the third from top curve in Fig.~\ref{szprychy}b,  
$\ppk$ keeps increasing up to a maximum value (larger than $90^\circ$ but
smaller than $135^\circ$), which is reached at the minimum
observed $V$ (i.e.~at the maximum amplitude of the negative $V$).

So far we can imagine that we moved horizontally across the oval 
centered at $(\psin, \lag)=(45^\circ, 90^\circ)$ in Fig.~\ref{snake}, 
and that we stopped near the tip of the wavy arrow. 
According to the third from top line in Fig.~\ref{szprychy}b, 
$\ppk$ is now decreasing towards the vicinity of the $90^\circ$ mode, 
i.e.~$V/I$ comes back to zero, and the distorted 
PA track (marked `b' in Fig.~\ref{proto}a) approaches the RVM curve 
of the secondary (patch) mode.
At this longitude the core emission appears to cease, consistently 
with the third from top case in Fig.~\ref{szprychy}b. Therefore, 
the average observed PA leaves the loop-shaped distortion and 
jumps to the primary mode of the peripheric
(``conal") emission. Thus, the mode identity replacement in B1237$+$25 
occurs only within the core component, and does not propagate 
onto the peripheric emission. The primary mode on both sides of the profile 
must then follow the same RVM track.

\begin{figure}
\includegraphics[width=0.48\textwidth]{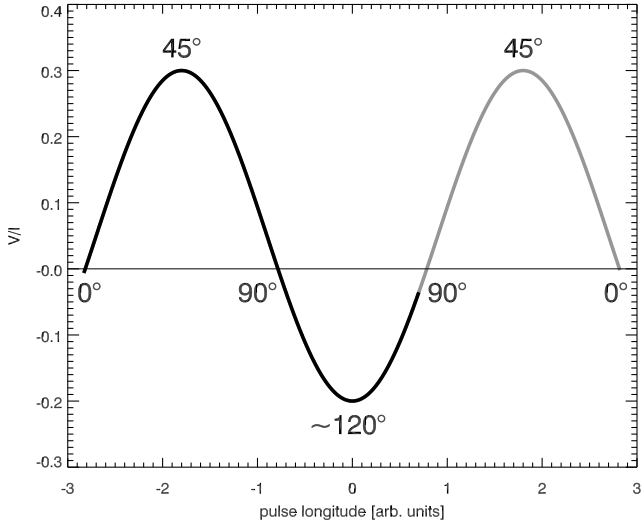}
\caption{Generic profile of core $V/I$ for the case when $\ppk$ exceeds
the upper mode value ($90^\circ$) in the center of a profile 
(e.g.~the third from top case in Fig.~\ref{szprychy}b).
The numbers give the values of $\ppk$. The full feature 
is symmetric around $\Phi=0$ (cf.~$V/I$ in B1541$+$09 and B1839$+$09). 
The misalignment of the PLR basis, however, 
locates the core beam on the left hand side, so only the approximately 
antisymmetric 
 black part of the $V/I$ profile may be observed (as in B1237$+$25). 
}
\label{vplot}
\end{figure}

A full derotation of $\ppk$ back to $0^\circ$ would 
produce a positive $V$ on the right-hand side of the core.
This would create a symmetric feature shown in Fig.~\ref{vplot},
with two humps of
positive $V$ separated by a negative $V$ in the middle. 
However, in B1237$+$25 the core beam is narrow in comparison to $\Delta_{plr}$
and the right hand side of the $\ppk(\Phi)$ profile is missing, 
as implied by Figs.~\ref{szprychy}b and \ref{szbis}b.  
Such interpretation is supported by the polarisation 
profiles of B1541$+$09 at 430 MHz, and B1839$+$09 at 1.4 GHz (see Figs.~8 and 9
in HR10) which exhibit the full symmetric $V$ profile, 
with the negative $V$ surrounded by the positive maxima of $V$.  

\subsubsection{The abnormal mode branch of the core loop in B1237$+$25}
\label{abnormal}

It is not clear if $\npsi$  
bifurcates on the left side of the core PA loop in B1237$+$25, 
because the bottom branch of the loop (marked `c' in
Fig.~\ref{proto}a) may represent the primary OPM track. 
Moreover, the `a-b' and 'c' branches of the loop dominate 
in different profile modes (normal N, and abnormal Ab, SRM13), 
i.e.~the bifurcation may be considered 
nonsimultaneous. The enhancement of the `c' branch in the Ab mode 
is associated with 
 a strong change of the profile shape: the fourth component 
(counting from left) disappears, 
and new emission appears between 
the second component and the core (see Fig.~6 in SRM13). 
In terms of the CMA model, the downward deflection of the 'c' branch 
in the Ab mode requires that $\npsi$ moves in the opposite direction 
than in the N mode, i.e.~towards the smaller $\ppk$.
According to Fig.~\ref{snake}, this should reverse the sign of $V$, which is
not observed. 
Therefore, the change of the $\npsi$ motion direction 
should be accompanied by a change of the sign of $\lpk$, i.e.~$\nlag$ should 
become displaced to the other side of $\lag=0$.\footnote{It is possible 
to interpret this phenomenon 
in terms of the spiral radio beam geometry (Dyks 2017). \nct{dyk17}
A change of sign of the electric field in the polar region would change the 
direction of the $\vec E\times\vec B$ drift, so the spiral would 
revolve in the opposite direction, while possibly being 
anchored at the same point. 
This could displace the emission of the 4th component to the space 
between the component no.~2 and the core. Simultaneously, charges of opposite
sign (than in the N mode) could be accelerated and emitting 
towards the observer. The change of the spiral geometry, as driven 
by the reversal of the electric field, would then produce the profile mode
change, or, in general, a change in drifting or fluctuating 
single pulse properties.} 

\subsection{Separation of profiles into OPMs}

The coherent nature of the observed non-RVM PA distortions implies 
that the incoherent separation of pulsar emission into OPMs, which is based on 
RVM fits, may not give meaningfull results in the presence of strong PA
distortions. Moreover, since the flux ratio 
of the OPMs mostly reflects the relative magnitude of components of 
the input polarisation vector (and the width of $\npsi$), 
the intensity profiles of separated OPMs tell little on the OPMs
themselves. McKinnon \& Stinebring (2000) provide other arguments 
against the modal separation of profiles.

\subsection{PA distortions caused by a complicated magnetic field}

The complicated loop-like distortions of the core PA 
are then a result of fast and multiple 
intermode crossing that occurs when the sightline is passing near the pole,
and the projected $B$ field rotates with respect to the 
PLR basis (which is also rotating). 
Similar fast variations of $\ppk$ may be expected when the magnetic field
in the radio emission region has a complicated multipolar structure 
(e.g.~Petri 2017). \nct{pet17}

The CMA model implies that such multipolar distortions of $B$ 
do not leave a direct
imprint of this local $B$ in the observed PA curves. As in the core case, 
the multipolar $B$ barely changes the illumination of the PLR basis, so
 the resulting PA distortions appear as transitions between 
(or departures from) the RVM tracks. These `transitional' distortions
are caused by the intermode crossing and reflect the relative
orientation of $B$ in the emission and PLR regions, not the 
absolute direction of $B$ in the multipolar emission region. The numerous 
distortions of a single PA curve observed in millisecond pulsars
(e.g.~J0437$-$4715) may be caused by such multipolar $B$.

\subsection{The $45^\circ$ PA jumps in weakly polarised profiles}
\label{sec45}

\begin{figure}
\includegraphics[width=0.48\textwidth]{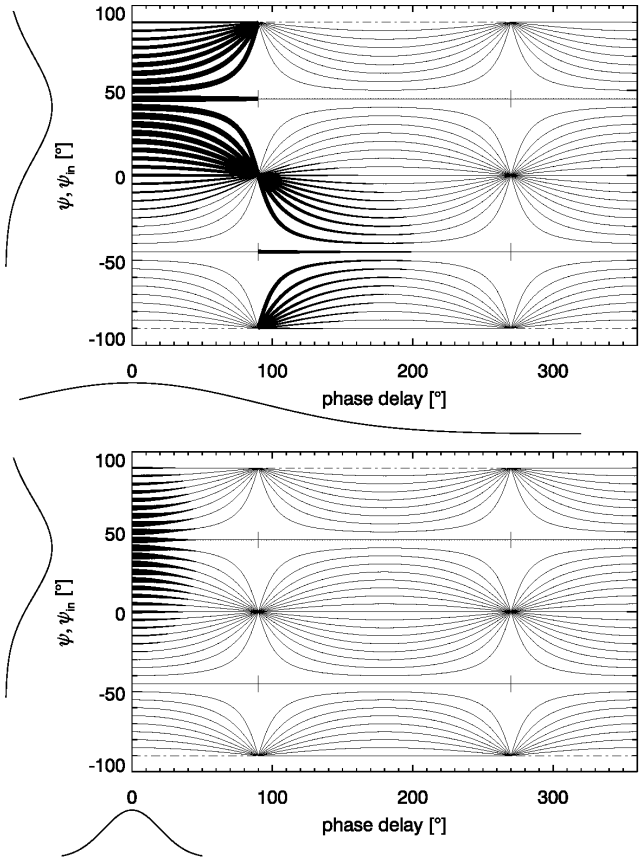}
\caption{Mechanism of the $45^\circ$ PA jump. The narrowing of 
$\nlag$ (shown below the panels) switches off the OPMs and makes 
the intrinsic (incident) PA distribution observable.
}
\label{45}
\end{figure}

A rare but striking polarisation effect is presented by 
pulsars B1919$+$21 (Fig.~18 in MAR15) and B0823$+$26 (Fig.~7 in 
Everett \& Weisberg
2001), 
\nct{ew2001}
in which the PA makes a $45^\circ$ jump and follows a continuous 
PA curve associated with a very weakly polarised emission. 

The low $L/I$ suggests that $\ppk$ is close to $45^\circ$ or that $\npsi$ is
very wide. The first case is shown in Fig.~\ref{45}, in which 
$\ppk=40^\circ$ and $\sigma_{\psi, in}=30^\circ$.
At the profile edges in B1919$+$21, $\nlag$ must be wide enough to 
reach the modes at $\lag=90^\circ$, as suggested by the orthogonal  
PA patches observed 
at $\Phi\approx-1.5$ in Fig.~18 in MAR15. 
This is presented in the top panel of Fig.~\ref{45}, 
where $\sigma_{\lag}=90^\circ$.
Since $\npsi$ is not perfectly aligned with $\ppk=45^\circ$,  
one of the OPM spots is slightly stronger, which defines which OPM is observed
as the average  PA. 
In the middle of the profile of B1919$+$21, 
$\nlag$ becomes narrower, as shown in  
the bottom panel of Fig.~\ref{45}. Therefore, the intrinsic (input) PA
distribution (also shown with the line thickness at $\lag=0$) 
dominates there. Since $\npsi$ peaks
near $45^\circ$, the average PA offset by $\sim\negthinspace45^\circ$ 
is observed,
and the single pulse samples reveal the variety of incident PAs 
in the intrinsic $\npsi$ distribution. The `dirty', erratic single pulse PA 
observed in the inner parts of B1919$+$21 profile must then represent 
the intrinsic (incident) PA, or at least be closer to the intrinsic PA than
in the profile edges where the OPMs dominate. 
Needless to say, the $45^\circ$ jumps also appear when $\nlag$ stays narrow,
 but moves from $\lag\sim0$ to $\sim\negthinspace90^\circ$ 
(as in Fig.~\ref{ortho3}).

The transition from the wide to narrow $\nlag$ should in general
be associated with a change of $L/I$ at the $45^\circ$ jump. 
This is because for the wide $\nlag$ (top in Fig.~\ref{45}) 
the two OPM spots of comparable strength produce nearly total depolarisation, 
whereas for the narrow $\nlag$ (bottom) $L/I$ has the intrinsic value, 
which can be large if $\npsi$ is narrow. The magnitude of $L/I$ 
observed on both sides of  
the $45^\circ$ jump in B0823$+$26 (Fig.~7 in Everett \& Weisberg 2001, 
$\Phi=-2^\circ$) is indeed very different. 
If, on the other hand, the incident $\npsi$ is wide (and centred close to
$45^\circ$), then $L/I$ 
is low on both sides of the $45^\circ$ jump. This must be the case   
of B1919$+$21 which seems to have low $L/I$ both at the edges and 
in the middle of profile.

This interpretation suggests that the 
widening of $\nlag$ in the profile peripheries contributes to the
`edge depolarisation' observed
in pulsar profiles (Rankin \& Ramachandran 2003). \nct{rr03}

\subsection{PA jumps of arbitrary magnitude}

When $\npsi$ is not close to $45^\circ$, the fast narrowing of $\nlag$ 
produces PA jumps of arbitrary magnitude. These should be associated with a
generally larger $L/I$ (than in the $45^\circ$ case).

\subsection{Frequency dependent profile depolarisation}

Since the refraction coefficient (hence the speed) 
of a modal wave likely depends 
on the frequency $\nu$, the phase lag distribution $\nlag$ 
in Fig.~\ref{modes} may be expected to 
move horizontally (or change width) with varying $\nu$. This may be related 
to the observed $\nu$-dependent profile depolarisation.
However, early ventures
 into the interpretation of the observed $\nu$-dependent effects, 
suggest that the distribution of $\npsi$ strongly influences the look 
of profile polarisation at different frequencies.
This subject is deferred to future work.

\section{Summary}

It has been found that the observed pulsar OPMs and the strong circular
polarisation, are the statistical result of coherent addition 
of waves in two natural propagation modes. Precisely, the modes appear 
because the distribution of phase lags of combining waves extends 
up to $\lag=90^\circ$.
The combining 
waves represent the 
components of a single incident signal, which at lower altitudes 
may well be in a pure linearly polarised state.
 The pulsar emission mechanism (such as the extraordinary mode curvature
radiation, see Dyks 2017) \nct{dyk17} 
may then emit a linearly polarised radiation which initially 
propagates in a single mode (Melrose 2003). 

The production of OPMs and $V$ is indeed a propagation effect.
The original (emitted) signal is just illuminating the 
polarisation basis of intervening matter at a higher altitude.
The observed RVM-like variations of PA mostly 
reflect the orientation of magnetic field in this high-altitude intervening
 region. 
The orientation of $B$-field in the emission region, on the other hand, 
barely determines the ratio 
of flux in both observed modes 
(the emitted radiation is illuminating the PLR 
 $B$-field structure from below).

The polarisation characteristics that result from such mode origin 
are completely different from the properties of incoherently summed 
elliptically polarised natural modes. In the coherent case, a vector split at the angle of
$\psin=45^\circ$ generates strong nonzero $V$ (and in general a nonzero $L/I$,
depending on the modal phase lag). The change of $V$ sign occurs  
at a mode maximum intensity (in general at large $L/I$). $L$ and $V$ are 
anticorellated (unlike in the incoherent case). As shown in previous
sections, these characteristics are consistent 
with numerous observed properties of pulsar
polarisation. However, the CMA does not exclude effects governed by 
the standard incoherent summation of elliptically polarised natural modes 
(the incoherent summation corresponds to the wide $\nlag$ cases 
discussed in this paper, but the ellipticity of the proper modes has
not yet been included). 

Despite the suggestive look of the antisymmetric $V$ at core components, 
(which in the incoherent case would imply an OPM transition), 
it has been found that the change of $V$ sign is instead associated with 
the alignment of incident $\ppk$ with one of the modes. 
 The minimum in $L/I$ 
is indeed observed to coincide with peaks in $|V|/I$, not with the sign change.
The generally low $L/I$ within the entire core 
has its origin in wide or multiply peaked $\npsi$ distribution.
The antisymmetry of $V/I$ 
is consistent with the model provided that 
the symmetry point of the PLR polarisation basis is displaced 
rightwards with respect to the core.
For a negligible displacement, the symmetric $V$ profile is expected, 
as indeed observed in some objects.

The characteristic symmetry of the observed PA pattern, with the
 patches of the secondary mode in the profile peripheries, 
can be mostly attributed to the broadening of the input PA distribution
in the profile peripheries.
Because of the `PLR basis illumination',
the ratio of power in the observed orthogonal modes, 
as defined by their apparent location at presumed RVM tracks,
 is inversed whenever $\ppk$ crosses the intermode at $\pm45^\circ$.
An odd number of crossings makes the impression that the identity of 
primary/secondary modes has been replaced. 
This
happens within the core of B1237$+$25, but at the vanishing 
trailing edge of the core, the 
 mode designation is set back to original, through the orthogonal 
jump to the dominating primary mode of the ``conal" emission.




There has been a tendency among pulsarists to interpret 
various kinks in PA curves as a result of adjacent components being emitted
from different altitudes, or in terms of longitudinal polar currents.
Many of those distortions may in fact result from coherent mode combination. 
For example, the tiny distortion of the average PA curve under the 
components $C_3$ and $C_4$ in J1024$-$0719 (Fig.~5 in Craig 2014, 
longitude interval 
$[-50^\circ,-20^\circ]$), is clearly coincident with an increase in $V$ 
\nct{cra2014} which suggests the coherent origin.

Incoherent separation of intensity profiles into OPMs, 
as defined by an RVM fit to the average PA data may not produce a correct
result in the presence of strong distortions, since the latter have coherent
origin. 

It has been shown that the empirical approach to the 
coherent mode addition is the missing ingredient
needed to
understand numerous polarisation properties of radio pulsars. 
With the coherency allowed, it is possible to 
comprehend polarisation effects that are beyond the reach 
of the RVM model with the incoherently added modes.


\section*{acknowledgements}
I thank Joel Weisberg for permission to reproduce the figure
showing B1913$+$16. I am grateful to Adam Frankowski for 
detailed comments on the manuscript and discussions. 
I appreciate discussions with Bronek Rudak.
This work was supported by 
the National Science Centre grant DEC-2011/02/A/ST9/00256.
\bibliographystyle{mn2e}

\bibliography{listofrefs2}


\end{document}